\documentclass [12pt]{article}
\usepackage{latexsym}
\usepackage{amsfonts}
\usepackage{acronym}
\newcommand{\be}{\begin{equation}} \newcommand{\ee}{\end{equation}}
\newcommand{\bea}{\begin{eqnarray}}
\newcommand{\eea}{\end{eqnarray}}
\makeatletter
\@addtoreset{equation}{section}

 \makeatother
\begin{document}
\normalsize
\title {Minkowskian Yang-Mills vacuum}
\author
{{\bf L.~D.~Lantsman}\\
 Wissenschaftliche Gesellschaft bei
 J$\ddot u$dische Gemeinde  zu Rostock,\\Augusten Strasse, 20,\\
 18055, Rostock, Germany; \\ 
Tel.  049-0381-799-07-24,\\
llantsman@freenet.de}
\medskip
\maketitle
\begin {abstract}
The well-known Bogomol'nyi-Prasad-Sommerfeld  (BPS) monopole is
considered in the limit of the infinite mass of the Higgs field as a basis for constructing
the Yang-Mills  vacuum with the finite energy density. In this limit the
Higgs field disappears at the spatial infinity, but it leaves, nevertheless,  its trace as vacuum  Yang-Mills  BPS monopoles transformed
into  Wu-Yang monopoles obtained  in the pure Yang-Mills theory by  a
spontaneous scale symmetry breaking in the class of functions with  zero
topological charges.
The topological degeneration of  a vacuum BPS monopole  manifests itself via 
Gribov copies of the covariant Coulomb gauge in the form of the time integral
of the Gauss law constraint. We also show that, in the considered  theory,  there is a zero mode
of the Gauss constraint involving an "electric" monopole and the additional mass
of the $\eta'$-meson in Minkowskian QCD. The consequences of the Minkowskian physical monopole vacuum:  rising "golden section" potential and
topological confinement, are studied in the framework of the perturbation
theory. An estimation of the vacuum expectation value of the square of the
magnetic tension is  given through the $\eta'$-meson mass, and arguments in
favour of the stability of the monopole vacuum are considered. We also discuss
why all these "smiles" of the Cheshire cat are kept by the Dirac fundamental
quantization, but not by the conventional Faddeev-Popov integral.
\end{abstract}
\noindent PACS:  14.80.Bn,  14.80.Hv     \newline
Keywords: Non-Abelian Theory, BPS Monopole, Minkowski Space,
Topological Degeneration, Wu-Yang Monopole, Infrared Topological Confinement.
\newpage

\section{Introduction.}
The nature of  the vacuum of the Yang-Mills (YM) theory
in the Minkowski space still remains an open problem at the present time. 
There were a lot of attempts to solve this problem. \par
A typical feature of  these attempts was constructing  the
nontrivial physical vacuum in the Minkowski space  on the basis of
nonzero values of field vacuum expectations,  coinciding with  classical fields. \par
As an  example of these attempts
we should like to point out the work \cite{Mat}
 stimulated by the asymptotic freedom
formula, as a criterion for instability of the naive perturbations
theory \cite{Gr}. \par
However, these attempts did not take  account of the
topological structure of vacuum.\par
This nontrivial topological structure of the YM vacuum was discovered  in the Euclidean space $E_4$ \cite{Bel}.  This implied 
that there exist classical \it in- \rm and out-\rm
 vacuum states corresponding to different topological indices
 $\vert n> $ with zero values of energy and
 tunnel transitions $\vert n>\to \vert n+1>$ ($n\in{\bf Z}$) 
 occur between them. These  transitions are described by \it instantons\rm,
 i.e.  YM fields with  fixed topological numbers
 $\nu= n_{ out}- n_{ in}$ on which  the YM action attains its minimum corresponding to the zero eigenvalue of energy.  \par 
The  defects of this vacuum are: the unphysical status of this zero value of energy
in quantum theory and  explicit violation of the Poincare (CP)  invariance. \par
However,  the topological degeneration of 
initial data for YM fields
does not depend on the space where these fields are considered.
The initial data of any classical solution in the Minkowski space-time
are also topologically degenerated. Therefore it is worth to investigate topologically
degenerated vacuum solutions in  the Minkowski space-time
in the class of functions with physical values of finite energy densities \rm.
 \par
In the present  paper we attempt to explain the advantages of  going over from the Euclidian  space-time to the Minkowski one in non-Abelian gauge theories.
 \par
 This work  is devoted to just such investigation of the
 nontrivial topological vacuum  inherent in  the YM theory (as a striking pattern of 
non-Abelian gauge models) 
in  the Minkowski space-time. \par
 The existence of such a vacuum is stipulated  by the fact that the homotopies groups of
all the three-dimensional  paths (loops) in the $SU(2)$  group manifold is
 \it nontrivial \rm (see p. 325 in  \cite{Al.S.}):
\be
\label{top1}
\pi_3 (SU(2)) =\bf Z.
\ee
 We always should     take account of this "initial" topology at the analysis of the YM vacuum
(independently on the space where we study it). \par
We investigate the topological degeneration of the initial
 data using the well-known \it Bogomol'nyi-Prasad-Sommerfeld \rm (BPS)
 monopole as an example. Such monopoles appear as a  result of the spontaneous
breakdown of the initial $SU(2)$ symmetry ($SU(2)\to U(1)$) in the presence of the Higgs isoscalar multiplet (generalizing  
 the classical Higgs  field
$\phi$ in the well-known  $\lambda \phi^4$ theory)  
in the limit  $\lambda\to 0$ for the Higgs selfinteraction. \par 
In the  lowest order by the temperature $T$ (effectively,  in a neighbourhood
of \linebreak $T\to 0$)
we may  always separate  in this  gas the Bose condensate (at temperatures below 
a critical  temperature $T_0\sim 0$)
and quantum excitations over
this Bose condensate.  
\par
Thus there is a possibility  to construct the YM vacuum using
the Bose condensate of free scalar particles in the limit
of their infinite masses,
\footnote{The difference between the Bose condensate of free scalar particles
and Minkowskian YM vacuum is, indeed, in the existence of the (Higgs-YM) interaction with the coupling constant $g$.
This turns the  (Higgs-YM) vacuum \it into the  c-number Bose condensate in a non-ideal Bose gas, in which  inevitable arise
long-range correlations of local excitations and
cooperative degrees of freedom\rm. 
 Such a system is alike to the superfluid helium II \cite{N.N.}.}
  when these particles disappear from the spectrum
of elementary excitations of the theory, leaving therein, nevertheless, their various "traces". The study of these "traces" is just the  goal
of the present paper.
\par
One of these "traces" is the topological degeneration of the
BPS monopole perturbation theory, 
 manifesting itself via  \it Gribov copies \rm of
the covariant Coulomb gauge, treated, in turn, as initial data of the Gauss law constraint 
in the lowest order of the perturbations theory involving the "new" Minkowskian monopole vacuum.
These  Gribov copies  imply that there exists a zero mode solution
to the Gauss law constraint  expressed through the \it global dynamical
variable \rm $N(t)$. This zero mode solution
correctly  describes the collective solid potential rotation of the  (YM-Higgs) Bose
condensate  with the real energy-momentum spectrum.
\par
We also construct the generating functional for weak perturbation excitations
over this vacuum in the form of the Feynman path integral.\par
The  ensuing exposition is organized as
follows.\par
In Section 2 we  show  that  in the Minkowskian YM theory, in the so-called BPS limit $\lambda \to 0$; $m\to 0$ \cite{BPS} for the Higgs mass and Higgs selfinteraction respectively, there exist 
 nontrivial BPS monopole
solutions to  the  equations of motion involving  finite energies densities.
They corresponds to the $SU(2)\to U(1)$ spontaneous breakdown. \par
The Bogomol'nyi equation \cite{BPS}, specifying the lowest level of the BPS monopole configuration energy,
sets  the immediate correlation between YM and Higgs multiplets belonging to the $U(1)\to SU(2)$ embedding. \par 
This will be a starting point for the construction of a consistent
 theory of
the \linebreak Minkowskian YM vacuum in  Section 4.\par
 Section 3 is devoted to
  constructing  the Dirac  variables in the general
YM theory. We demonstrate that they   take the look of solutions to the Gauss
 constraint-shell equation. This will be the base of all our
 further discussion in the present work. \par
The topological degeneration of the Minkowskian YM initial data is the subject of
Section 4. \par
We  argue, in Subsection 4.1,  in favour of that the vacuum in the "old" instanton
approach \cite{Bel} is, indeed,  not the physical one\rm. \par 
As an alternative, in Subsections 4.2 and 4.3, we construct  "Minkowskian" 
vacuum  monopoles $\Phi_i^{(0)}({\bf x})$  in the form of a
stationary Bose condensate involving  topological numbers  $n=0$ and the nonzero vacuum 
"magnetic" tension $B(\Phi_i^{(0)})$ corresponding these YM monopoles. \par
 All this is a result
of the $SU(2)\to U(1)$ spontaneous breakdown, describing  by  the classical equations
of the Minkowskian non-Abelian theory in the class of fields with  topological
numbers $n=0$. \par
These equations permit  nontrivial solutions, at the spatial infinity, in the form of
 Wu-Yang monopoles \cite{Wu}: $\Phi_i^{(0)}({\bf x})$. \par
Present constructing  the
Minkowskian YM vacuum is only a presentation of such solutions  as BPS monopoles
 in the theory involving Higgs fields in the BPS limit \cite{BPS} of their infinite masses (associated with the infinite spatial volume $V$),
  but at finite energies densities. \par
 Herewith
 the vacuum "magnetic" tension $B(\Phi_i^{(0)})$, specified by the Bogomol'nyi equation  \cite{BPS}, 
  itself acquires a crucial importance.\par 
 We show that in the considered limit the  Gibbs expectation value
 $<B^2>$ (specified by means averaging $B^2$ over the spatial volume)
 is, indeed, different from zero; in this is an 
 analogy  with the \it Meisner effect \rm in a superconductor. \par
The value $<B^2>\neq 0$, inherent in the Minkowskian YM model involving vacuum BPS monopoles, is,  in turn, a precondition to solving  the $\eta '$-meson problem in Minkowskian QCD. 
\par
The nonzero value of
$<B^2>$ allows  us also  to regularize our theory by  introducing  an
infrared cut-off parameter $\epsilon(<B^2>)$  playing the role of the typical 
size of  BPS monopoles.\par
The principal goal of Section 4 is to reveal the nature of the topological
degeneration of  the YM vacuum  monopole $\Phi_i^{(0)}({\bf x})$ and the close correlation between this topological degeneration and the Gribov
 ambiguity in the choose of the covariant Coulomb  gauge  (in the form of the
 time integral of the Gauss law constraint). \par
This topological  degeneration
 is determined by  non-perturbation multipliers \linebreak
 $\exp(n \hat\Phi_0({\bf x}))$, with $\hat \Phi_0({\bf x})$ being a solution
 to the Gribov ambiguity equation having the look of a Higgs BPS monopole.\par
In Subsection 4.3 we quote the explicit expression  \cite{LP1} for these gauge
transformations, turned YM fields into \it topological \rm Dirac variables satisfied the covariant Coulomb gauge and herewith gauge invariant.  
\par  
As far as the covariant Coulomb gauge is, indeed, the time integral of the Gauss law
 constraint, the Gribov ambiguity signals us that there is the  zero
 mode solution  to the Gauss law constraint, treated as the equation to the temporal
 component $A_0^c(t,{\bf x})$ of a YM field.\par
The in the main new step in our investigations about the Minkowskian YM theory is  introducing, in Subsection 4.4,   the continuous topological variable $N(t)$ for  specifying  the zero mode
of the Gauss law constraint. This allows us to represent any Minkowskian YM field 
$A_0^c(t,{\bf x})$ as the product
$\dot N(t)\Phi_0^{(0)c}({\bf x})$. \par  
 This zero mode solution induces the vacuum
"electric" tension ("\it electric \rm" \it monopole\rm) as a dynamic degree of freedom that
cannot  be removed by fixing  any gauge. \par
This "electric" tension,
 in turn, generates  the free rotator action 
describing the collective solid potential rotation of the Minkowskian (YM-Higgs) vacuum. 
\par
The appropriate
Schr\"odinger equation for this vacuum gives  the real energy-momentum spectrum, unlike  
the one in the instanton YM theory \cite{Bel}. \par 
The dependence of the free rotator action  on the vacuum expectation value
 $<B^2>$, which, in  turn, depends on the Higgs mass
 (through the Bogomol'nyi equation), confirms our assumption about
the Minkowskian YM vacuum as a Bose condensate.\par
The topic of Section 5 is a more detailed analysis of  the zero  mode solution to the Gauss law constraint 
and the YM (constraint-shell)
action; also we decompose the YM "electric" tension into its
transverse and  longitudinal parts with respect to  the constraint-shell (Gauss) 
equation. We investigate consequences of this decomposing.\par
 Section 6 is devoted to the calculation of the instantaneous potential of
 the current-current interaction in the presence of  Wu-Yang background monopoles \cite{Wu}. \par 
  The
   YM Green function in the Wu-Yang monopoles background takes, indeed, the shape  of the sum of 
 two potentials. There are
 the Coulomb potential and the rising "golden section"
 one.  \par 
This result has a great importance for the  analysis of the
 hadronization and $\eta '$-meson problem.\par
The analysis of the  Feynman and FP path integrals is the subject of
Section 7.\par
The last two  Sections, 8 and 9, are devoted to the analysis of the infrared 
topological confinement, involving  the quark
confinement in QCD as direct consequences of the average over
the topological degeneration. \par
The theory considered in Sections 8 and 9  allow us to assert that only the
colourless ("hadronic") states form a complete set of physical
states in (Minkowskian) QCD. We  prove that
the infrared 
topological confinement implies  the quarks  confinement in  (Minkowskian) QCD, that
 the complete set of  hadronic  states
ensures that this QCD is, indeed, a unitary theory. \par 
 In Section 10 we estimate the value of
  the vacuum chromomagnetic field in QCD${}_{(3+1)}$ and point out the way to solve the $U(1)$-problem in the  Minkowskian YM theory involving vacuum BPS monopoles.
\section{ Gauge Higgs effect in  Minkowski space.}
Our idea, basing onto our  discussion in the previous subsection, is to construct the physical Minkowskian 
  YM vacuum using the Higgs Bose condensate in the Minkowskian YM 
theories involving monopoles \cite{BPS,Hooft, Polyakov}. \par
Herewith we desire to utilize   the well-known 
\it Bogomol'nyi-Prasad-Sommerfeld \rm (BPS) \it  limit
 of the zero self-interaction\rm: $\lambda \to 0$ (at $m \to 0$)  in  the Higgs sector of the Minkowskian 
YM action  (see, e.g., \cite{Al.S.,Gold})
\be
\label{YM L}
S=-\frac {1}{4 g^2} \int d^4x F_{\mu \nu}^b F_b^{\mu \nu }+
\frac {1}{2} \int d^4x (D_\mu\phi,D^\mu\phi
) -\frac {\lambda}{4} \int d^4x \left[(\phi^b)^2- \frac{m^2}{\lambda}\right]^2,
\ee
with 
$$D_\mu\phi=\partial^\mu\phi+g[A^{\mu },\phi]$$
 being the YM
covariant derivative and $g$ being the YM coupling constant.\par
We suppose that  the initial data of all the  fields are given to within stationary
gauge transformations, the  manifold of these transformations has  a
nontrivial structure of   three-dimensional paths in the group space of
the (initial) non-Abelian $SU(2)$ gauge group:
\be
\label{3-path}
\pi_3 (SU(2)) =\bf Z,
\ee
with  $\bf Z$ being  the group of integers: $n=0,\pm 1,.\pm 2,..$.
\par
In the case of  the $SU(2)$ gauge theory
the Yang-Mills fields $A^{\mu b}$ and Higgs fields $\phi^b$ take
their values in the Lie algebra  of the  $SU(2)$ group.\par
To obtain the converging action integral, corresponding to  finite values of energy, we should claim  for the Higgs
field 
$\phi ({\bf r})$ to be finite as ${\bf r} \to \infty$ in 
the  BPS limit \rm  $\lambda\to 0$.\par
This implies that $\phi^a$ would go to
the minimum of the potential $V\equiv \frac {\lambda}{4} (\frac {m^2}{\lambda}-\phi ^2)^2$:
\be
\label{Higs.as}
\phi^{a \infty} ({\bf n})\in M_0, \quad {\bf n} =\frac {\bf r}{r},
\ee
where $M_0$ is the  manifold of the minimum of the potential $V$(\it the vacuum
manifold \rm ):
\be
\label{min}
M_0=\{\phi= a;\quad a^2=m^2/ \lambda\}
\ee
as ${\bf r} \to \infty$. Thus $M_0$ consists of the points of the two-sphere
$S^2$ in the three-dimensional $SU(2)$  group space.\par 
The presence of Higgs \it pseudo-Goldstone \rm modes
\cite{Weinberg} implies that
the initial $SU(2)$ gauge symmetry inherent in the (Minkowskian) YM model is then 
 spontaneously violated down to  its $U(1)$ subgroup (via the Higgs mechanism 
of the  $SU(2)$ gauge symmetry breakdown:
see, e.g., pp. 243- 244 in  \cite{Cheng}).
\par
On the face of it, we may choose the Higgs isovector  $\vec \phi $ along the axis $z$ in
 the Cartesian coordinates:
\be
\label{phi}
\vec \phi =(0,0,m/\sqrt{\lambda}),
\ee
as a  ground state configuration. Thus this vector  stays
invariant under rotations around the axis $z$ ($U(1)$
transformations).
\par
Note, however, that the choice (\ref{phi}) in the whole space is, indeed,
\it topologically trivial\rm. Really (see \cite {Al.S.}, \S $\Phi$4), the gauge condition
$$\phi _i=0, \quad i=1,2; \quad 
\phi_3=\vert \vec \phi\vert $$ 
is not compatible with nontrivial
topologies $n\neq 0$. For the field satisfying such gauge condition
its asymptotic at the spatial infinity
is trivial: the solutions of the  look
\be
\label{2.6} \vec\phi^{\infty} ({\bf n})= {\bf V} ({\bf n})\vec \phi,\ee 
with $V ({\bf n})$ being a
continuous function of $\bf n$ (${\bf n} = {\bf r}/r $ is the unit radius of the spatial sphere $S^2$)
  taking its values in the  $SU(2)$  group space 
 in the case of the YM theory, are topologically equivalent to
$\vec \phi =(0,0,a)$. On the other hand, ${\bf V} ({\bf n})$, considered as the map $\pi _2 (SU(2))$,
is equal to zero.
\par
 We should define the topological structure of the
manifold (\ref{min}). \par
In
the case of a discrete group  $G$,  $\phi^{a \infty}$ would be
constant,  as long as it is a continuous function
(from the topological point of view, we deal  in this case
with the group $\pi_0$ \cite{Switz} of 
connection components, that is trivial in the case of a
connected manifold; the sphere $S^2:=\{{\bf n} =1$ as $ {\bf r} \to \infty \}$ is
 an example of
 such manifolds).
In this case $\phi^{a \infty}$ has a trivial topology. \par
If  ${\rm dim} ( M_0) \neq 0$, $ M_0$ has a nontrivial topology.
Therefore the group of symmetry $G$ \linebreak \it would be continuous\rm.
One may be shown (see, e.g. pp. 465- 466 in \cite{Cheng}) that the covariant
derivatives $D_i\phi $, entering the action  (\ref{YM L}), decreases as $r^{-2}$;
thus  the integral (\ref{YM L}) is, indeed, finite. This guarantees  nontrivial topological
features of the Minkowskian YM theory.\par
Issuing from Eq. (\ref{min}), that specifies the manifold $M_0$ of the
minimum of the potential $V$, 
we see that the sphere $S^2\simeq M_0$
maps into the sphere $S^2:=\{{\bf n} =1\}$ as
 $ {\bf r} \to \infty$.
This map has the nontrivial homotopies group of two-dimensional loops:
\be
\label{top2}
\pi_2 S^2= \pi_3 (SU(2))=\pi_1(U(1))=\pi_1~S^1=\bf Z. \ee 
Just this nontrivial topology determines \it  magnetic charges \rm
associated  with the residual $U(1)$ gauge symmetry (these charges alone point to
an "electromagnetic" theory). The presence of  magnetic charges  implies 
that there exists the solution to the  equations of motion 
for  the YM action  (\ref{YM L}) \it in the class of
magnetic monopoles\rm, i.e. \it the stationary vacuum solutions at the spatial  infinity
corresponding to the quantum-field configuration of the
minimum energy\rm, $E_{min}$ (according to our definition of the
vacuum as a ground state of the minimum energy).
We may  write down this monopole solution.\par
In particular, the Higgs isovector would be directly proportional to  $\bf n$
as $ {\bf r} \to \infty$: in the light of the said above   about the map
$S^2\simeq M_0 \to
S^2:=\{{\bf n} =1\}$ as  $ {\bf r} \to \infty$. Thus its look would be
\be
\label{hedg}
\phi^a \sim \frac {x^a}{r}f(r,a)
\ee
as $ {\bf r} \to \infty$; $f(r,a)$ is a continuous function  that does
not change the topology (\ref{top2}).\par
This solution for $\phi^a$ appears for the first time in  the work \cite {Polyakov},
and it is called  the \it hedgehog\rm. A good analysis
of hedgehogs also is  carried out   in the monograph \cite {Linde} (pp. 
114-   116). \par
There may be shown (see \S $\Phi$11 in \cite {Al.S.}) that there
 exists the (vacuum)  solution to the  equations of motion (in the zero topological sector of the Minkowskian YM theory),
\it regular in a finite spatial
volume \rm  and generalizing Polyakov hedgehogs (\ref {hedg})  
\footnote{ The statement that the said 
solution is regular in a finite spatial
volume  implies that we should consider
the topology (\ref{top2}) and the vacuum manifold $M_0$,
(\ref{min}), taking account of this finite spatial
volume. If we wish to adapt our theory to the needs of (Minkowskian)
QCD (we shall see how to do this in Sections 8 and 9), the spatial
volume  specified by the typical hadronic
size, $\sim 1$ fm. ($\sim 5$ GeV$^{-1}$), is quite sufficient for
our  purposes.}, in the form
\cite {Al.S.,Gold}
\be
\label{sc monopol}
\phi^a =
 \frac{ x^a}{gr} f_0^{BPS}(r)~,~~~~~~~~~~~~~~~~~~~~~~~~~~~~~
 f_0^{BPS}(r)=\left[  \frac{1}{\epsilon\tanh(r/\epsilon)}-\frac{1}{r}\right],
\ee
\be
\label{YM monopol}
 A^a_i(t,\vec x)\equiv\Phi^{aBPS}_i(\vec x) =\epsilon_{iak}\frac{x^k}{gr^2}f^{BPS}_{1}(r),~~~~~~~~
  f^{BPS}_{1}= \left[1 -
 \frac{r}{\epsilon \sinh(r/\epsilon)}\right],
\ee  
obtained in the  BPS limit
\be
\label{lim}
\lambda\to 0,~~~~~~m\to 0:~~~~~~~~~~
~~~~~\frac{1}{\epsilon}\equiv\frac{gm}{\sqrt{\lambda}}\not =0.
\ee
The Higgs Bose condensate behaves as the Bose condensate in an ideal gas in   this   limit. \par
The functions $f_0^{BPS}(r)$ and   $f^{BPS}_{1} (r)$ are called 
the \it BPS ansatzes\rm, while the solutions (\ref {sc monopol}), (\ref {YM monopol}) are called \bf BPS monopoles\rm.  \par
The vacuum  solution (\ref {sc monopol}), (\ref {YM monopol}) satisfies the \it potentiality condition\rm: 
\be
\label{Bog}
{\bf B} =\pm D\vec \phi,
\ee
with $\bf B$ being the vacuum "magnetic" tension in the theory (\ref{YM L}).
This equation (called \it  the Bogomol'nyi equation\rm) is obtained
at  evaluating
 the lowest bound of the (YM-Higgs) energy:
\be
\label{Emin}
E_{min}= 4\pi {\bf m }\frac {a}{g},~~~~~~~~~~~~\,\,
~~~~~a=\frac{m}{\sqrt{\lambda}}
\ee
 (with $\bf m$ being the magnetic charge), for the BPS monopole solutions.\par
The Bogomol'nyi  equation (\ref{Bog}) shows  that there exists
a nonzero vacuum  "magnetic"   tension ${\bf B}$ in the Minkowskian YM theory involving vacuum BPS monopoles  in its Higgs and YM sectors\rm. \par 
The Bogomol'nyi  equation (\ref{Bog}) may be rewritten in the tensor
form  as \cite {Al.S.}
\be
\label{Bog1} \frac {1}{2g}\epsilon ^{ijk}F_{jk} =\nabla ^i\phi.
\ee 
We see that the Bogomol'nyi equation (\ref{Bog}) allows us to get a
consistent theory involving   YM and Higgs  $U(1)\to SU(2)$ multiplets and
yielding the solutions of the BPS monopole type.
The Higgs sector of that theory defines the $U(1)$ group of gauge symmetry
with the nontrivial
topology (\ref{top2}), involving  magnetic charges and   radial "magnetic" (vacuum) fields. \par
The said may serve as a  good base for  constructing the consistent  theory
for the  physical \rm YM vacuum in the Minkowski space.\par
In contrast to the "old" Euclidian approach \cite{Bel} to  the YM vacuum,
our "Minkowskian" conception of the  YM vacuum as a stationary Bose condensate with the simultaneous strong (YM-Higgs) coupling \cite{Pervush1}
yields, indeed, the
real spectrum of momentum. \par
We shall  make sure  that the stationary vacuum
fields in the new Minkowskian YM theory possess the winding numbers $n=0$ and
they indeed have the look of BPS  monopoles, (\ref {sc monopol}), (\ref {YM monopol}) respectively. \par
The "electric" and "magnetic" tensions  corresponding to these vacuum fields also 
will  be constructed. \par
The topological degeneration (for $n\neq 0$) in the theory here represented is
realized via  \it Gribov copies \rm of  the covariant Coulomb
gauge imposed
on  (vacuum) YM potentials. \par
The YM (gluonic) fields are
considered as  weak perturbation excitations (\it multipoles\rm )
over this  BPS monopole vacuum.
These
excitations  have the asymptotic $O(\frac {1}{r^{1+l}}),~l>1$ at the
 spatial infinity. \par
All this will be discussed in the next sections. 
\section{Dirac quantization of  Yang-Mills theory.}
Let us consider the "pure"  YM theory  involving the   gauge $SU(2)$ group
in the four-\linebreak dimensional Minkowski space-time. The action of that theory
is given by the formula
\be
\label{act YM}
W[ A _\mu]=-\frac {1}{4} \int d^4x F_{\mu \nu}^a F_a^{\mu \nu } 
 = \frac {1}{2}\int d^4x (F_{oi}^{a2}- B_i^{a2}),\ee
where the standard definitions of the non-Abelian "electric" tension $F_{oi}^{a}$: \be
\label{electr}
F_{0i}^a = \partial _0 A^a_i- D(A)^{ab}_i A_{0b}, \quad  D^{ab}_i = (\delta ^{ab}\partial_i+ g \epsilon ^{acb}A_{ci}),
\ee
and the "magnetic" one, $B_i^a$:
\be
\label{magnet}
B_i^a= \epsilon_{ijk} (\partial^j A^{ak} +\frac {g}{2}\epsilon ^{abc}A_{b}^j A_{c}^k),
\ee
are used. The action (\ref{act YM}) is invariant with respect to the gauge transformations $u(t; {\bf x})$:
\be
\label{gauge}
{\hat A}^u_i=u(t; {\bf x})({\hat A}_i+\partial_i)u^{-1}(t,{\bf x}),\ee
with ${\hat A}_\mu= g \frac {\tau ^a}{2i}A_{a\mu}$.\par
 Solutions to the non-Abelian constraint equation
(\it  the Gauss law constraint \rm ):
\be
\label{Gauss}
\frac {\delta W}{\delta A^a_0}=0,\Longleftrightarrow [D^2(A)]^{ac}A_{0c}= D^{ac}_i(A)\partial_0 A_{c}^i,
\ee
and to the  equation of motion:
\be
\label{mot}
\frac {\delta W}{\delta A^a_i}=0,\Longleftrightarrow [\delta_{ij}D^2_k(A)-D_j(A)D_i(A)]^{ac}A_{c}^j= D^{ac}_0(A)[\partial^0 A_{ci}- D(A)_{cbi} A^{ob}],
\ee
are specified by  boundary conditions and  initial data.
They  generalize  the appropriate  equations
in the  Maxwell electrodynamics
(see Eqs. (7), (8) in \cite{Pervush2}). \par
 The Gauss law constraint  (\ref{Gauss})  associates  initial data of $A^0_a$
 to the one of  the spatial components $A^i_a$.\par
To remove the unphysical variables, we may solve this constraint in the form of the naive perturbation
series:
\be
\label{ser}
A^0_c= \frac {1}{\Delta}\partial_0 \partial_iA_{c}^i +\dots,
\ee
with $\Delta$  being the Laplacian. As we remember from mathematical physics (see, e.g., p. 203 in
\cite {Vlad}),
the \it fundamental
solution \rm
to the \it Laplace equation\rm:
\be
\label{Laplace}
\Delta {\cal E}_3 =\delta (x),
\ee
is
\be
\label{fund}
{\cal E}_3 = -\frac {1}{4\pi x}.
\ee
This  specifies the action of the operator $\Delta^{-1}$ on
a continuous function $f(x)$:
\be
\label{Delta}
\Delta^{-1}f(x)=- \frac {1}{4\pi } \int d^3 y\frac {f(y)}{\vert x-y\vert},
\ee
with $\Delta^{-1}$   being  the \it Coulomb kernel of the non-local distribution \rm
(see also (12) in \cite {Pervush2}).\par
Thus  resolving  the (YM) Gauss law constraint and  substituting  this
solution into the equations of motion distinguishes  
gauge invariant non-local (radiation) variables. \par
Upon  substituting  this solution into  Eq.
(\ref{mot}) the lowest order of this equation in the
coupling constant $g$ contains \it only  transverse  fields \rm
(this level mathematically coincides, as a \it linearized YM
theory\rm, with the theory of  radiation variables in QED \cite {Pervush2}):
\be
\label{QED}
[\partial_0^2-\Delta] A^{cT}_k+\dots =0 , \quad A^{cT}_i= [\delta_{ik}- \partial_i\Delta^{-1}\partial_k]A^{ck} +\dots 
\ee
This perturbation theory is well known as  the \it radiation \rm \cite{Schwinger}
or \it  Coulomb \rm  \cite{Fadd1,Gitman} gauge
 involving the generating functional of Green functions in the
form of a Feynman integral in the rest
  reference frame $l^{(0)}=(1,0,0,0)$:
\bea
\label{Fein}
Z_F[l^{(0)},J^{aT}]&=&\int\limits_{ }^{ }
 \int\limits_{ }^{ }\prod\limits_{c=1 }^{c=3}
 [d^2A^{cT} d^2 E_{cT}]\nonumber\\
 &&\times\exp\left\{iW^T_{l^{(0)}} [A^T,E^T]-i\int\limits_{ }^{
 }d^4x [J^{cT}_{k} \cdot A_{cT}^{k}]\right\},
\eea 
with the constraint-shell action:
\be
\label{csha}
W_{l^(0)}^T[A^T,E^T] = W^I\vert \sb {\frac{\delta W^I}{\delta A_0} =0},
\ee
given in the  first order formalism (see \cite{Slavnov}, p. 83):
\be
\label{WI}
W^I = \int d t\int d^3 x \{ F^c_{0i}E_c^i- \frac {1}{2}[E^c_i E^i_c+ B^c_i B^i_c]\}=\int d t\int d^3 x (E^{ci}\partial_0 A_{ci}+A_{0c}D^c-
H),
\ee
where
\be
\label{der}
D^c=\partial_k E^{kc}-g[A_k^b,E^{kd}]\epsilon _{bd}^c,
\ee
and
\be
\label{Ham}
H=\frac {1}{2} (E_k^{c2}+B_k^{c2})=\frac {1}{2}[(E^{Tc})^2+(\partial_i \sigma ^c)^2+B_k^{c2}]
\ee
is the Hamiltonian of the YM theory. \par
We decompose here the "electric" tension $ E^{kc}$ \it into the
transverse and longitudinal parts \rm respectively:
\be
\label{decomp}
E^c_i= E^{Tc}_i+ \partial_i \sigma ^c, \quad \partial_i E^{Tc}_i= 0 .
\ee
The constraint
\be
\label{constr1}
\frac{\delta W^I}{\delta A_0}=0 \Longleftrightarrow D^{cd}_i (A)E_d^i=0
\ee
may be solved in terms of \it radiation \rm  variables. \par
The function $\sigma ^a$ has the look \cite {Fadd1}
\be
\label{sigma}
\sigma ^a[A^T,E^T]= (\frac {1}{D_i (A)\partial^i})^{ac}\epsilon_{cbd}A_k^{Tb}E^{Tkd}\equiv (\frac {1}{\hat \Delta})^{ac}
\epsilon_{cbd}A_k^{Tb}E^{Tkd}.
\ee
It is worth to notice that the vector $D^c$ in (\ref{der}), 
that is the nothing else as the contrvariant derivative of the "electric" field $E_i^c$, disappears, indeed, on the surface of the Gauss law constraint (\ref{constr1}).
\par
Note also (see Eq. (16.24) in \cite {Gitman}) that  ${\rm det} [D_i(A)\partial^i]$ in (\ref{sigma})
is precisely \it the Faddeev-Popov \rm (FP) \it determinant \rm in the YM Hamiltonian formalism corresponding to the transverse (Coulomb) gauge of YM fields:
\be
\label{FP}
\hat \Delta ^b_a A_{0b}-\partial_i E^i_a=0,\quad \hat \Delta ^b_a \equiv D ^b_{ai}\partial^i,
\ee
with
\be
\label{moment}
E_{ia}= \frac {\partial {\cal L}}{\partial \dot A^{ia}}=F_{0i}^a
\ee
being the canonical momentum (\ref{electr}).\par
A complete proof that  ${\rm det}\hat \Delta ^b_a$ is, indeed, the FP determinant of the YM theory is given in  the monograph \cite {Gitman}, where   there was shown that the  radiation gauge
in the YM theory is equivalent to the FP determinant
 ${\rm det}\hat \Delta ^b_a$ (see (16.30) in \cite {Gitman}).\par
The operator quantization of  the YM theory in terms of the
radiation variables belongs to Schwinger \cite {Schwinger},
who proved the relativistic covariance of  the radiation variables (\ref{QED}). \par
This  implies that the radiation  fields are transformed as
 non-local functionals (\it Dirac  variables \rm \cite {Pervush2}),
\be
\label{Dir.v}
{\hat A}_k^T[A]= v^T[A]({\hat A}_k+\partial_k) (v^T[A])^{-1},\quad {\hat A}_k^T = 
g\frac {A_k^{Ta}\tau _a}{2i},
\ee
where  matrices $v^T[A]$ are  specified issuing from  the  claim to YM fields to be 
transverse: $\partial_k A^{kT}=0$.\par
These  matrices also would  cancel the action of the $SU(2)$ gauge group,
  (\ref{gauge}), onto YM fields, to ensure the gauge invariance  of Dirac  variables. More precisely, matrices $v^T[A]$ would be transformed as
\be
\label{z-n dlja v}
v^T[A]\to v^T_u [A]= u^{-1}(t,{\bf x})~ v^T[A]
\ee
at  the gauge transformations (\ref{gauge}) \cite{David2}.
\par
It is less obviously that the Dirac  variables (\ref{Dir.v}) are actually transverse. 
A good analysis of  this fact was carried out in the review \cite{Pervush2} it is the linearized YM theory
(\ref{ser}), (\ref{QED}). \par
One may be shown, utilizing the arguments of the Dirac  variables  analysis in QED 
\cite{Pervush2} (see also   \cite{LP1, Pervush1, David2}),
that the Dirac  variables (\ref{Dir.v}) are transverse if and only if the Dirac matrices $v^T[A]$ have the look
\be
\label{v} v^T(t,{\bf x})= v^T({\bf x})T \exp \{\int  \limits_{t_0}^t d {\bar t}\hat A _0(\bar t, 
{\bf x})\}; \quad v^T(t,{\bf x})\vert_{t=t_0}= v^T({\bf x})
\ee
(the symbol $T$ denotes the time ordering of the matrices under the exponent sign),
where, resolving the Gauss law constraint  (\ref{Gauss}), we represent the temporal component of an YM field, $\hat A _0$, as the
series (\ref{ser}) in the linearized YM theory, i.e. as a non-local functional of YM fields.
Just this proves that the Dirac  variables (\ref{Dir.v}) are indeed transverse. \par
Our discussion about the Dirac matrices $v^T[A]$
we shall continue in Section 4.3. We shall show that the
topological degeneration of initial (Minkowskian) YM data is determined only by stationary matrices 
$v^T({\bf x})$. The latter one comes to Gribov topological factors \cite{Gribov} $v^{(n)}({\bf  x})\equiv \exp(n \hat\Phi_0({\bf  x}))$, with $n\in {\bf Z}$ and $\hat\Phi_0({\bf  x})$ being the so-called Gribov phase, whose explicit look we shall ascertain in Section 4.3 \cite{LP1,Pervush2,David1}. Ibid we also shall generalize the linearized YM theory
(\ref{ser}) and  rewrite the exponential multiplier in (\ref{cl.vac}) in the form of the so-called "Dirac dressing" \cite{LP1,Pervush2} of non-Abelian fields, depending explicitly on 
Minkowskian vacuum  YM solutions: BPS and Wu-Yang monopoles. \par
Note that the removal \cite{Dir} of temporal components of  YM fields  (contradicting the Dirac quantization principles) implies that the appropriate "temporal" Dirac  variables are equal to zero \cite{David2}: 
\be \label{udalenie} v^T[A](A_0(t)+\partial_0)(v^{T})^{-1}[A]=0.
\ee
We may treat the latter formula as an equation for specifying   
Dirac matrices $v^T[A]$. \par 
At the level of the Feynman path integral
the relativistic covariance implies a relativistic transformation of sources
(this, in turn, involves    the appropriate relativistic transformations of     the Dirac variables (\ref{Dir.v})  \cite {Pervush2}).\par
The definition (\ref{Dir.v}) of the Dirac variables may be interpreted as
a transition to  new variables, allowing  us to
rewrite the Feynman integral (\ref {Fein}) in the form of the appropriate FP integral \cite{Fadd1,Fadd2,Fradkin1}:
\bea
\label{ymfpi}
 Z_F[l^{(0)},J^{aT}]&=&\int\limits_{ }^{ }
 \int\limits_{ }^{ }\prod\limits_{c=1 }^{c=3 }
 [d^4A^{c} ]\delta(\partial_i A^{ci} ){\rm Det}[D_i(A)\partial^i] \nonumber\\
 &&\times\exp\left\{i W [A]-i\int\limits_{ }^{
 }d^4x (J^{T}_{ck} \cdot A^{Tkc}[A])\right\}~.
 \eea
It was proved in \cite {Fadd1,Fadd2,Fradkin1} that, \it on  mass-shells of
 radiation  fields, the scattering amplitudes do not depend on  factors
\rm $ v^T[A]$. But the following question is quite reasonable:
\it why we cannot
 observe these scattering amplitudes\rm?
There are a few answers to this question: the infrared instability of the
 naive perturbation theory \cite{Mat,Nils},
the Gribov ambiguity, or the zero value of the FP determinant \cite {Gribov},
the topological degeneration of the physical states \cite{Pervush3,Ilieva,Nguyen}. \par
This will be the subject of our discussion in the next sections.
\section{Topological degeneration of initial data.}
\subsection{Instanton theory.}
One may  find a lot of solutions to equations of classical Maxwell electrodynamics.
 Nature chooses, however,  two types of  functions:
 \it monopoles \rm (electric charges), that specify
non-local electrostatic phenomena (including instantaneous bound states)
 and
 \it multipoles\rm, that specify
spatial components of  gauge  fields involving nonzero magnetic tensions.\par
Spatial components of  non-Abelian  fields, considered above as the
radiation variables (\ref{QED}) in the naive perturbation theory
(\ref{ser}), are also defined as multipoles. \par
In the non-Abelian theory, however, it is a reason, as we have seen this in
Section 2, to assume that \it
 spatial components of  non-Abelian  fields\rm: for example, YM vacuum BPS monopoles (\ref{YM monopol}), \it belong to the
monopole class of functions\rm, like  temporal components of  Abelian
fields (as, for instance, the Coulomb potential),  that we 
also treat as  monopoles.
\par
This fact was revealed by the authors of  the instanton YM theory \cite {Bel}.
Instantons satisfy the duality equation in the Euclidean space $E_4$
(where the Hodge duality operator $*$ has the $\pm 1$ eigenvalues for
 external 2-forms defining the YM tension tensor);
thus the instanton YM action coincides, in effect, with  the \it Chern-Simons functional
\rm
(the \it Pontryagin index\rm) (see, e.g., Eq. (10.104) in \cite {Ryder}):
\be
\label{Ch-S}
\nu [A]=\frac {g^2}{16\pi ^2}\int\limits_{t_{in}}^{t_{out}}
dt\int d^3 x F_{\mu \nu}^a {{}^*F}^{\mu \nu}_a = X[A_{out}]-X[A_{in}]=
n(t_{out})-n(t_{in}),
\ee
with (see (10.93) in \cite {Ryder})
\be
\label{wind}
X[A]=-\frac {1}{8\pi ^2}\int{\sb V} d^3 x \epsilon ^{ijk}
{\rm tr} [{\hat A}_i \partial_j{\hat A}_k- \frac {2}{3}{\hat A}_i{\hat A}_j
{\hat A}_k];\quad A_{in,out}= A(t_{in,out},x);
\ee
being the \it topological winding number functional of  gauge  fields \rm
 and $n\in \bf Z $ being the value of this functional for the classical vacuum:
\be
\label{cl.vac}
{\hat A}_i= L^n_i= v^{(n)}({\bf x})\partial_i v^{(n)}({\bf x})^{-1}.
\ee
The manifold of all the classical vacua  in a non-Abelian theory
represents \it the group of  three-dimensional paths \rm lying in the
three-dimensional $SU(2)$-manifold with the   homotopies group
$\pi_3 (SU(2))= {\bf Z}$. \par
The whole group of stationary matrices $v^{(n)}({\bf x})$ is split into the
topological classes marked by the  integer topological numbers
(the \it  Pontryagin degrees of the map\rm)
specified by the expression (see (10.106) in \cite {Ryder}) 
\be
\label{degree}
{\cal N}[n]=- \frac {1}{24\pi ^2}\int d^3 x \epsilon ^{ijk} {\rm tr} [L^n_iL^n_jL^n_k],
\ee
that shows \cite {LP1} how many times a three-dimensional path $v^{(n)} ({\bf x})$
 turns around the $SU(2)$ group manifold when the coordinate $x_i$
runs over the space where it is specified.\par
Gribov, in 1976, proposed to consider   instantons as
\it  Euclidean solutions interpolating between  classical vacua with different
degrees of the map \rm (or as \it tunnel transitions between these classical vacua\rm).\par
The degree of the map (\ref{degree}) may be treated as a normalization condition  that specifies the class of functions for the given
classical vacuum (\ref{cl.vac}). \par
In particular, to obtain  Eq. (\ref{cl.vac}),
we should choose the classical vacuum in the form
\be
\label{Ins.deg}
v^{(n)}({\bf x})=\exp (n {\hat \Phi}_0({\bf x})),
\quad {\hat \Phi}_0 =- i\pi \frac {\tau ^a x_a}{r} f_0 (r)\quad (r= \vert {\bf r}
\vert )
\ee
(compare with (16.34) in \cite{Cheng}; we should also set $x_0=0$ in this formula
for  stationary gauge transformations).\par
The function $f_0 (r)$ satisfies the boundary conditions 
\be
\label{cond}
f_0 (0)=0,\quad f_0 (\infty)=1.
\ee
Note a  parallel between this solution and Eq. (\ref{hedg}).\par
The common between the Minkowskian monopole and Euclidian instanton YM theories is, indeed, \it in
the similar  topological structure inherent in the both theories\rm.\par
In the case of  the YM  instanton theory \cite{Bel} we deal  with the map (\ref{3-path}):
$S^3\to SU(2)$ as
${\bf x} \to \infty$. This induces the homotopies group
$\pi_3(SU (2))= \pi_3 S^3 ={\bf Z}$ (where $ S^3$ is the bound of the Euclidian space
$E_4$)
coinciding with $\pi_2 S^2=\bf Z$ (see (\ref{top2}))
in the Minkowskian theory (\ref{YM L})- (\ref{Bog}), involving vacuum BPS monopoles in the YM and Higgs sectors of that theory. \par
This generates similar theories. \par
 But there exists also the principal
distinction between  the both theories. \par
As a consequence of the  relation 
$$\pi_3(SU (2))= \pi_3 S^3 ={\bf Z},$$
instantons \bf may exist in the Euclidian YM theory \cite{Bel} 
without any spontaneous $SU(2)$ breakdown\rm. 
This breakdown is not a necessary thing in this case,
and
we consider $SU(2)$ as an \it exact symmetry \rm in the instanton YM
theory \cite{Bel}.\par
 Thus we obtain the  solution of the monopole type
in (\ref{cl.vac}) as ${\bf x} \to \infty$.\par
To show that these classical values are not sufficient
to describe  a physical vacuum in the non-Abelian (Euclidian) theory, we should consider
a \it quantum instanton\rm, i.e. a corresponding to the zero vacuum solution to
the Schr\"odinger equation \cite{LP1,Pervush1}
\be
\label{Schrod}
{\hat H} \Psi _0[A]=0,
\ee
with
${\hat H}= \int d^3x [E^2+B^2]$, $E= \frac {\delta}{i\delta A }$
being the operators of the Hamiltonian and field momentum respectively.
This solution may be constructed by using the winding number functional (\ref{wind})
and its variation derivative,
\be
\label{d.wind}
\frac {\delta}{\delta A^c_i} X[A]=\frac {g^2}{16\pi ^2} B^c_i(A).
\ee
The vacuum wave functional, rewriting in terms of the winding number functional
(\ref{wind}), has the look \it of a  plane wave  \rm \cite {Pervush1}:
\be
\label{plan}
\Psi _0[A]=\exp (iP_{\cal N}X[A]),
\ee
for \it unphysical \rm  imaginary values of the topological
momentum $P_{\cal N}=\pm 8\pi i/g^2$ \cite {Pervush1,Arsen}.\par
We should like to note that in QED
 this type of the wave functional belongs to the unphysical part of the
spectrum, like the wave function of the harmonious oscillator
 $({\hat p}^2+q^2)\phi_0=0$.
The value of this unphysical plane wave functional
\footnote{The wave function (\ref {plan}) is not good normalized, the
imaginary 
topological
momentum $P_{\cal N}=\pm 8\pi i/g^2$ turns it in a
function with the non-integrable square.}
for the  classical
vacuum (\ref{cl.vac}) coincides with the \it semi-classical instanton wave function
\rm
\be
\label{wave}
\exp (iW [A_{instanton}]=\Psi _0[A=L_{out}]\times \Psi^* _0[A=L_{in}]=
\exp(-\frac {8\pi ^2}{g^2}[n_{out}-n_{in}]).
\ee
This exact relation between a semi-classical instanton and its quantum
version (\ref{Schrod}) points out that  classical YM instantons \cite{Bel} 
are also
unphysical solutions\rm; they permanently tunnel  in the
Euclidean space-time $E_4$ between the classical vacua with  zero energies
that do not belong to the physical spectrum.
\subsection{Physical vacuum and  gauge Higgs effect.}
Our next  step is the assertion \cite{Jack} about the topological
degeneration of  initial data not only  of  the classical vacuum but also of
all the physical  fields with respect to the stationary gauge transformations
\be
\label{stat}
{\hat A}_i^{(n)}(t_0,{\bf x})= v^{(n)}({\bf x}){\hat A}_i^{(0)}(t_0,{\bf x})v^{(n)}({\bf x})^{-1}+L^n_i,
\quad L^n_i=
v^{(n)}({\bf x}) \partial _i v^{(n)}({\bf x})^{-1}.
\ee The stationary transformations $ v^{(n)}({\bf x})$ with $n = 0$ are
 called \it the small one\rm; and those with $n\neq0$ are
 called \it the
large ones \rm  \cite {Jack}.\par
The group of transformations (\ref{stat}) implies that \it
 spatial components of  non-Abelian  fields, involving  nonzero magnetic
 tensions \rm $B(A)\neq 0$ \it belong to the monopole class of
functions\rm,  like  temporal components of  Abelian  fields.
In this case  non-Abelian  fields   involving nonzero magnetic tensions
contain  non-perturbation monopole-type terms, and  spatial components
may be decomposed \cite {Pervush1} into  sums of 
vacuum monopoles
$\Phi _i ^{(0)}({\bf x})$ and  multipoles ${\bar A} _i$:
\be
\label{sum}
A_i ^{(0)}(t_0,{\bf x})=\Phi _i ^{(0)}({\bf x})+ {\bar {\hat A}}_i ^{(0)}(t_0,{\bf x}).
\ee
Each multipole is treated as a weak perturbation over the appropriate monopole having the following 
asymptotic at the  spatial infinity \cite {LP1}:
\be
\label{asym}
{\bar  A}_i (t_0,{\bf x})\vert _{\rm asymptotic}= O (\frac {1}{r^{l+1}})\quad (l>1).
\ee
Nielsen and Olesen \cite {Nils} and Matinyan and Savidy \cite {Mat}
introduced the  vacuum "magnetic" tension using the fact that all the
asymptotically free theories are instable, and the perturbation vacuum is
not the lowest stable state.\par
The extension of the topological classification of  classical vacua to all the
initial data of the spatial components helps us to choose 
vacuum monopoles with  the zero value of the winding number functional (\ref{Ch-S}):
\be
\label{choose}
X[A=\Phi_i^{c(0)}]=0,\quad \frac {\delta X[A]}{\delta A_i ^c} \vert _{A=\Phi^{(0)}} \neq 0.
\ee
The zero value of the winding number functional, the claim for non-Abelian fields to be transverse, and the spherical
symmetry of  vacuum monopoles fix the class of initial data for  spatial
components:
\be
\label{vec}
{\hat\Phi}_i =-i \frac {\tau ^a}{2}\epsilon _{iak}\frac {x^k}{r^2}f(r).
\ee
They contain an unknown function $f(r)$. The classical equation for this
function takes the form \cite {LP1}
\be
\label{eq}
D^{ab}_k(\Phi_i)F^{bk}_a(\Phi_i)=0 \Longrightarrow \frac {d^2f}{d r^2}+\frac {f (f^2-1)}{r^2}=0.
\ee
One  may see  three solutions to this equation:
\be
\label{3}
f_1^{PT}=0,\quad f_1^{WY}=\pm 1\quad (r\neq 0).
\ee
The  first solution corresponds \it to the naive instable perturbation
theory, involving the asymptotic freedom formula \rm \cite {Mat,Gr}. \par
Two nontrivial solutions $ f_1^{WY}=\pm 1$ are also well known.
They are \it the Wu-Yang monopoles \rm \cite {Wu}, applied to construct  physical variables
in the work \cite{Niemi} (\it the hedgehog and the antihedgehog\rm, respectively, in the
terminology  \cite{Linde,Polyakov}). \par
As it was shown in the papers \cite{LP1,Pervush2,David2,David1} (and we shall repeat these arguments in Section 6),
 Wu-Yang monopoles involve the rising potential of the instantaneous interaction between two
total \cite{David1} excitation currents. This interaction
rearranges the perturbation series, leads to the to the appearance of the gluonic constituent mass and removes the asymptotic freedom
formula \cite{Bogolubskaja,Werner}
as an origin of   instability in the non-Abelian model considered.\par
The appearance of said two 
 solutions with the opposite signs
in (\ref{3}) is very remarkable, especially  as we approximate
our BPS monopole solutions by  Wu-Yang monopole ones; it is 
easy to see that we obtain then the BPS ansatzes $\pm f_1^{BPS}(r) $ in  (\ref{YM monopol}) instead of $f_1^{WY}$ in (\ref{3}). 
\par
Wu-Yang monopoles \cite{Wu},  thus, are solutions to the classical equations everywhere
\it excepting an infinite small
neighbourhood of
 the origin of coordinates\rm, $r = 0$. The
appropriate "magnetic" field is
\be
\label{B}
B_i ^a(\Phi_k)= \frac {x^ax^i}{gr^4}.
\ee
To remove this singularity of  Wu-Yang monopoles \cite{Wu} at the origin of coordinates and to regularize its
energy,  Wu-Yang monopoles \cite{Wu} are considered as the limit (as $r\to \infty$) of YM
BPS monopoles (\ref{YM monopol}):
\be
\label{ansatz}
f_1^{BPS}=[1-
\frac {r}{\epsilon \sinh  {r/\epsilon}}]\Longrightarrow f_1^{WY},
\ee
when the mass of the Higgs field goes to infinity in the limit 
of the infinite spatial volume $V\to \infty$:
\be
\label{mass}
 \frac{1}{\epsilon}=\frac{gm}{\sqrt{\lambda}}=\frac{g^2<B^2>V}{4\pi}\to \infty.
\ee
Herewith  BPS monopoles result the finite energy density \cite{David2}:
\be
\label{magn.e}
\int \limits_{\epsilon}^{\infty } d^3x [B_i ^a(\Phi_k)]^2 \equiv V <B^2> 
= 4\pi \frac{gm}{g^2\sqrt{\lambda}}=\frac {4\pi} {g^2\epsilon} 
\equiv\frac{1}{\alpha_s\epsilon} .
\ee
The infra-red cut-off parameter  $\epsilon$ 
disappears in the limit $V \to \infty$, i.e. when the 
mass of the Higgs field goes to infinity and  Wu-Yang monopole
turn, in a continuous wise, into  BPS monopoles.  
In this case the BPS-regularization of  Wu-Yang monopoles
is similar to the \it infrared regularization in QED
\rm by  introducing 
the so-called "\it photon mass\rm" $\lambda$ (see,  e.g., \cite {A.I.}, p. 413) that also violates the initial
equations of motion
\footnote {It is necessary to note here that,  in quantum-field theories (QFT),  the transition to the limit $V \to \infty$
is usually performed upon  calculating physical  observable values:  scattering sections,
probabilities of decays and so on; herewith these values would be  
normalized per   time and volume units. Therefore all the specific features of the Minkowskian YM theory involving vacuum BPS   monopoles,
 including the topological degeneration  of initial data and vacuum "electric" monopoles (us discussed below), survive in any  finite spatial volume.  }.\par
The vacuum  energy density corresponding to   (BPS, Wu-Yang) monopole solutions:
\be
\label{dens}
\sim <B^2> \equiv \frac {1}{\alpha_s\epsilon V}=
\frac{4\pi m}{g\sqrt{\lambda} V},
\ee
may be removed by the  appropriate  counter-term
in the YM Lagrangian \cite {David2}:
\be
\label{contr}
{\bar{\cal L}}= {\cal L}-\frac {< B ^2 >}{2}.
\ee
In the existence of the nonzero vacuum "magnetic" tension $\bf B$, induced by the Bogomol'nyi  equation (\ref{Bog}),
is the crucial distinction of the topological
degeneration of  fields in  the Minkowski space from the topological
degeneration of the classical vacuum in the  instanton YM theory \cite{Bel} in the
Euclidean space $E_4$ (where the $F^{\mu \nu}(x)\to 0$ as $x\to \infty$ normalization of the Euclidian YM tension tensor 
is a precondition of Eq. (\ref {Ins.deg}): see pp. 482- 483 in \cite{Cheng}
\footnote{The author intends to give the detailed comparative analysis of the instanton Euclidian YM theory \cite{Bel} and Minkowskian YM theory involving vacuum BPS (Wu-Yang) monopole modes in the series of next their works. }).
The Bogol'nyi equation (\ref {Bog}), compatible with the topology
(\ref{top2}),  ensures the existence of this 
nonzero vacuum "magnetic" tension.
\par
The problem is, indeed, \it to formulate the Dirac quantization of  weak perturbations
of   non-Abelian  fields in the presence of 
non-perturbation monopole background\rm, taking  account of  the topological
 degeneration of all the initial data.\par
This will be the topic of the next subsection.
\subsection{ Dirac method and Gribov copies.}
Instead of the artificial equations of the  gauge-fixation method \cite{Fadd3}:
\be
\label{fix}
F(A_\mu)=0,\quad F(A_\mu ^u)= M_F u \neq 0 \Longrightarrow Z^{FP}= 
 \int \prod \sb {\mu} D A_\mu det M_F \delta (F(A))e^{iW},
\ee
we now repeat the Dirac \it constraint-shell \rm formulation resolving the Gauss law 
constraint (\ref{Gauss}) with  nonzero initial data \footnote{ Indeed 
(cf. Eqs. (15.12), (15.13) in \cite{Gitman}), the Gauss law constraint in
the YM theory would contain the additional  term, described \it the total  current \rm \cite{David1}. This is the sum of  
two items: the non-Abelian  and fermionic currents. In the Minkowskian YM model  here represented we treat this total current as a perturbation
over the topologically degenerated BPS monopole vacuum \cite{LP1,David1}. The consequences of this assumption we shall discuss in Sections 5 and 6.}:
\be
\label{init}
\partial _0 A_i ^c =0 \Longrightarrow  A_i ^c(t,{\bf x})=\Phi_i^{c(0)}({\bf x}).
\ee
The vacuum \it magneto-static \rm  field $\Phi_i^{c(0)}$ corresponds to the zero value
of the winding number (\ref{wind}), $X[\Phi_i^{c(0)}]=0$,
and satisfies the classical equations everywhere  excepting  a
small region near the origin of coordinates of the size 
\be
\label{size}
\epsilon \sim \frac {1}{\int d^3x  B ^2 (\Phi)} \equiv \frac {1}{< B ^2 >V},
\ee
(\ref{dens}), that disappears in the infinite volume limit.\par
The BPS monopole solutions for the Minkowskian YM vacuum in the zero topological sector, (\ref{YM monopol}), is one of examples
of  resolving (\ref{init}) the Gauss law constraint (\ref{Gauss}).\par
The second step is the consideration of  the perturbation theory (\ref{sum}), in which 
 the constraint (\ref{Gauss}) acquires  the look
\be
\label{Gauss1}
[D^2 (\Phi ^{(0)})]^{ac}A_{0c}^{(0)}=\partial _0 [D^{ac}_i(\Phi ^{(0)})A_c^{i ~(0)}].
\ee
Dirac proposed \cite{Dir} to remove YM temporal components $A_0$. \par
The quantization of these  non-dynamical
degrees of freedom contradicts the quantum principles. More precisely, the non-dynamic status of $A_0$ is not compatible with the
 quantization of this component due to its definite fixation  (e.g., through the Gauss law constraint), while  the appropriate zero canonical momentum
$$E_0=\partial {\cal L}/\partial (\partial_0
 A_0)=0$$
contradicts  the commutation relations and uncertainty principle  \cite{Pervush2}. \par  
 Thus the constraint (\ref{Gauss1}) acquires the form
\be
\label{Gauss2}
\partial _0 [D^{ac}_i(\Phi ^{(0)})A_c^{i~ (0)}]=0.
\ee
We define  the \it constraint-shell gauge \rm
\be
\label{cshg}
[D^{ac}_i(\Phi ^{(0)})A_c^{i ~(0)}]=0
\ee
as  zero initial data for this constraint. \par
It is easy to see that the expression in the square  brackets in
(\ref{cshg}) may be treated as \it the equal to zero, in the
initial  time instant $t=t_0$, longitudinal component of a stationary YM field \rm
(\ref {init}). We shall denote it as $A^{a\parallel}$:
\be
\label{Aparallel}
A^{a\parallel}\equiv [D^{ac}_i(\Phi ^{(0)})A_c^{i ~(0)}]=0\vert _{t=0}~.
\ee
Let us call the latter (\it Cauchy\rm) condition as  the 
\it covariant Coulomb gauge\rm.
Then the constraint (\ref{Gauss1})
 implies that also temporal derivatives of longitudinal
fields equal to zero. 
\par
The topological degeneration of initial data (with $n=0$) implies  \it that not only the
classical vacua, but also all the fields\rm, e.g.,   
\be \label{multipoli} A_i^{ (0)}=\Phi_i^{ (0)}+
{\bar A}_i^{ (0)},
\ee
 in the transverse gauge \rm (\ref{cshg})  are degenerated\rm:
\be
\label{degeneration1}
{\hat A}^{(n)}_i= v^{(n)}({\bf x}) ({\hat A}_i^{ (0)}+
\partial _i)v^{(n)}({\bf x})^{-1},\quad v^{(n)}({\bf x})=
\exp [n\Phi _0({\bf x})].
\ee
By analogy with (\ref{multipoli}), we may write down in any nonzero topological sector of the considered theory \cite {LP1}:
\be
\label{s1}
\hat A_i^{ (n)}(t,{\bf x}) =\Phi_i^{ (n)}({\bf x}) + {\hat{\bar A}}_i^{ (n)} (t,{\bf x}).
\ee
We may explicitly write down   gauge transformations leading to this topological degeneration
\cite {LP1}:
\bea
\label{degeneration}
\hat A_k = v^{(n)}({\bf x})T \exp \left\{\int  \limits_{t_0}^t d {\bar t}\hat A _0(\bar t,  {\bf x})\right\}\left({\hat A}_k^{(0)}+\partial_k\right ) \left[v^{(n)}({\bf x}) T \exp \left\{\int  \limits_{t_0}^t d {\bar t} \hat A _0(\bar t,{\bf x})\right\}\right]^{-1}, \eea 
where  the symbol $T$  stands for the  time ordering of the matrices under the exponent sign. \par
Thus in the initial time instant $t_0$ the topological degeneration of initial data
comes  to  "large" stationary matrices $v^{(n)}({\bf x})$, depending on  topological numbers $n\neq 0$ and called the factors of the
Gribov topological degeneration \cite{Gribov} or simply the \it Gribov factors\rm.  Thus we again come  to the matrices $v^T({\bf x})$ in (\ref{v}).
\par
The linearized YM  theory  (\ref{ser}),  (\ref{QED}) now prompts us  how to generalize
the exponential multipliers in   (\ref{degeneration}) for the Minkowskian YM theory involving the physical BPS monopole vacuum to obtain transverse and physical 
Dirac variables (satisfying the Coulomb gauge)
in such a model.
As far as we consider YM fields as perturbation excitations (multipoles)
over the physical BPS monopole vacuum (see (\ref{sum}), (\ref{multipoli}),
(\ref {s1})), it is obviously that the exponential multipliers in  
(\ref{degeneration}) would explicitly depend
on  YM BPS monopole modes (\ref{YM monopol}) (belonging to the zero topological sector),
turning, in a continuous wise, into Wu-Yang monopoles (\ref{vec})  at the spatial infinity.\par
Thus we come to the following look for the exponential multipliers in  
(\ref{degeneration}) 
\cite{LP1,Pervush1,Pervush2}:
\be \label{dress}
U(t,{\bf x})= v({\bf x})T \exp \{\int  \limits_{t_0}^t 
[\frac {1}{D^2(\Phi^{BPS })} \partial_0 D_k (\Phi^{BPS }){\hat A}^k]~d\bar t ~\}. \ee
Following the work  \cite{LP1}, we shall denote the exponential expression in (\ref{dress}) as $U^D[A]$; this expression may be rewritten \cite{LP1} as
\be
\label{UD}
U^D[A]= \exp \{\frac {1}{D^2(\Phi^{BPS })} D_k (\Phi^{BPS }) {\hat A}^k\}.
\ee
We shall call the matrices $U^D[A]$ as   \it Dirac \rm "\it dressing\rm" of non-Abelian fields. 
\footnote{In  the work  \cite{LP1} the Dirac dressing matrices $U^D[A]$ were represented in terms of Wu-Yang monopoles instead of BPS ones: (see, e.g.,   Eq. (C.1) in \cite{LP1}). It is, of course, quite correct, as BPS monopoles turn, in a continuous wise, into Wu-Yang monopoles. }
\par
The gauge transformations (\ref{degeneration}) are chosen to turn YM fields into
physical (transverse)  topological  Dirac variables \rm \cite{David2,David1}. \par 
Upon the transformation (\ref{cl.vac}),
the winding number functional (\ref{wind}) takes the  look (see Eq. (181) in \cite {Pervush2})
\be
\label{chang}
X[A^{(n)}_i]= X[A^{(0)}_i]+{\cal N}(n)+\frac {1}{8\pi ^2}\int d^3x \epsilon^{ijk}
 {\rm tr} [\partial _i({\hat A}_j^{ (0)} L^n_k)],
\ee
where the degree of the map ${\cal N}(n)=n$ ($n \in {\bf Z}$) is given by  Eq. (\ref{degree}).\par
The constraint-shell gauge (\ref{cshg}), (\ref{Aparallel}) keeps its look in each topological class:
\be
\label{transv}
D_i^{ab} (\Phi _k^{(n)}){\bar A}^{i(n)}_b =0, 
\ee
if the  phase $\hat\Phi _0({\bf x})$ of Gribov topological factors $v^{(n)}({\bf x})$ satisfies
\it the equation of the Gribov ambiguity \rm
\be
\label{Gribov.eq}
[D^2 _i(\Phi _k^{(0)})]^{ab}\Phi_{(0)b} =0;
\ee
this, in turn, implies  the zero FP determinant $\rm det$ $\hat \Delta$ in (\ref{FP}).\par
On the face of it, this zero FP determinant may involve nontrivial Dirac dressing matrices (\ref{UD}) when  we choose the Coulomb gauge  (\ref{Aparallel}). But in the  initial time instant $t=t_0$ the integral in (\ref{dress}) becomes zero; thus \it in this time instant the Gribov ambiguity equation \rm (\ref{Gribov.eq}) \it does not affect the gauge
transformations \rm (\ref{degeneration}),  i.e. the nature of topological Dirac variables and the Gribov topological degeneration\rm. 
\par 
In other words, \it in an arbitrary time instant \rm $t$ \it we may pick out a space-like
surface \rm ${\cal H}(t)$ \it in the Minkowski space-time where the topological degeneration
of initial YM data occurs\rm. This space-like surface is specified  by the  Gribov ambiguity  equation  (\ref{Gribov.eq}). 
\par
Note that \it the Gribov equation \rm (\ref{Gribov.eq}),
written down in terms of the Higgs isoscalar $\Phi_{(0)b}$ (it is nothing else as a topologically trivial vacuum Higgs BPS monopole mode (\ref{sc monopol})),
is the direct consequence of the Bogomol'nyj equation rewritten in the form (\ref{Bog1}) and 
Bianchi identity 
$$\epsilon ^{ijk}\nabla _i F_{jk}^b =0.$$
Note  that  Eqs.  
(\ref{Gauss2}), (\ref{Aparallel})  may be treated as \it  Cauchy  conditions \rm to the 
Gribov ambiguity equation (\ref{Gribov.eq}). \par
 Therefore to specify  the  
space-like surface  ${\cal H}(t_0)$
where the topological degeneration of initial YM data occurs, 
we should solve the  Cauchy  task (\ref{Gribov.eq})
with the  initial conditions (\ref{Gauss2}), (\ref{Aparallel}), i.e.  in the class of vacuum YM BPS monopoles
(\ref{init}) (and observable YM fields as  perturbation
excitations over this monopole vacuum possessing the same topological numbers  that
appropriate monopoles)  satisfying the Coulomb gauge  (\ref{Aparallel}) when we resolve the Gauss law constraint  (\ref{Gauss})  removing a la Dirac \cite{Dir} temporal components of YM fields.
\par
The existence of Gribov ambiguities in specifying  
transverse YM fields (satisfying the constraint-shell gauge 
 (\ref{Aparallel}), (\ref{transv})) in each topological class of the Minkowskian YM theory is a purely non-Abelian effect (see, e.g., \S T26 in \cite{Al.S.}). Just the ambiguity in the choice
of the Coulomb gauge (\ref{Aparallel}) is given by Eq. (\ref{Gribov.eq}) of the second order. \par 
The Gribov ambiguity equation (\ref{Gribov.eq}) permits  a very interesting geometric interpretation. \par 
We should recall, to begin with, that (see, e.g., \S T22 in \cite {Al.S.})
each gauge
field $A_\mu$, as an element of the adjoin representation of the given Lie
algebra, sets  an element $b_\gamma \in G$, where $G$ is the considered gauge
Lie group. These elements $b_\gamma$ are defined as
\be
\label{bg}
b_\gamma =P\exp (-\int {\sb \Gamma}  {\bf T} \cdot A _\mu dx^\mu),
\ee 
with $P$ being  the symbol of the parallel transfer along the curve $\gamma$
in a coordinate (for example, the Minkowski) space; $\bf T$ being the matrices of the
adjoin representation of the Lie algebra. \par
It is obviously that
$b_\gamma =b_{\gamma_1} b_{\gamma_2}$ as the end of the curve
$\Gamma_1$ coincides with the beginning of the curve $\Gamma_2$
and the curve $\Gamma$ is formed from these curves ($\Gamma=\Gamma_1\bigcup \Gamma_2$). Thus the group
operation (the multiplication) is associative; also always there exists 
the unit element and  element inverse to the given one.
This follows from the usual features of curves and the exponential 
function. \par
Thus elements $ b_\gamma $ form a subgroup in the given gauge group $G$.
Moreover, it is easy to see that this subgroup coincides with $G$ and may be treated as its specific (adjoin) representation. 
\par 
We see that elements $b_\gamma$ are specified over the set 
of  external 1-forms in the given Lie algebra. The cohomologies classes
of these external 1-forms are the elements of the cohomologies group $H^1$
(see, e.g., \S T7 in \cite {Al.S.}). \par
The Pontryagin formula
for a degree of a map (see, e.g.,  Lecture 26 in\cite{Postn3}):
\be
\label{deg}
\int \sb {X} f^*\omega = {\rm deg} ~f \in{\bf Z}  
\ee
(where the map $f:X\to Y$ is smooth and maps 
the compact space $X$ into the compact space $Y$: it is the 
so-called \it eigen map\rm; $f^*$ is the conjugate 
homomorphism of the appropriate cohomologies groups: $H^1_X\to H^1_Y$,
induced by the map $f$), sets a \it one-to-one correspondence
between the homotopies and cohomologies groups  \rm in the considered \linebreak theory
\footnote{We should apply Eq. (\ref {deg}) to the topology
(\ref{top2}) in the considered Minkowskian YM model.}. \par In particular,  
topological charges $n=0$ correspond to \it exact \rm 1-forms, i.e. to those
that may be represented as  differentials, $d\sigma$, of 
1-forms $\sigma$. Due to the \it Poincare lemma\rm,
$$d\cdot d\sigma =0,$$ 
i.e. \it each exact form is closed\rm.\par
A scalar field $\Phi_{(0)b}$  entering the
Gribov equation (\ref {Gribov.eq}) belongs to the zero topological sector: $n=0$. This zero 
charge, through the \it smooth \rm (and without additional singularity sources) 
 Bogomol'nyj equation (\ref {Bog1}) \cite{Al.S.}
\footnote{It is smooth, in effect, out of the $\epsilon$-neighbourhood of the origin of coordinates, where the value $\epsilon$ was specified by Eq. 
(\ref {mass}) as  the typical size of BPS monopoles \cite{Al.S.,LP1,David2,David1}. At the origin of coordinates the vacuum "magnetic" tension  $\bf B$ becomes infinite and diverges as $r^{-2}$ \cite{BPS}.}, is told to the
magnetic tension tensor $F^b_{jk}$ (i.e. to appropriate 2-forms) and to  
 appropriate YM fields (1-forms).\par
Let us now consider  those curves $\Gamma$ that begin and finish at
the same point $x_1$ of the Minkowski space. Such closed curves
are called \it the  one-dimensional cycles\rm, and the appropriate elements $b_\gamma$ may be written down as  cyclic integrals: 
\be
\label{cycl}
b_\gamma =P\exp (-\oint _{\Sigma} {\bf T} \cdot A _\mu dx^\mu),
\ee
over one-dimensional cycles $\Sigma$. According to the
\it De Rham theorem \rm (see p. 276 in \cite{Al.S.}), if
an external form $\omega$ is exact, its integral
over \it each \rm cycle specified over the considered manifold $M$
is equal to zero (this also confirms Eq.  (\ref{deg}) for topological charges $n=0$). \par
The said  implies that the 
integral in (\ref{cycl}) is equal to zero ($b_\gamma =1$) \it for each exact form \rm 
(corresponding to the zero topological charge due to (\ref{deg})).\par
The elements (\ref{cycl}) form \it the holonomies subgroup \rm $H$ in the given 
gauge group. In the considered Minkowskian YM  model \cite{LP1, Pervush2,David2,David1} it is the residual $U(1)$ gauge symmetry group, that is embedded in a nontrivial wise in the initial gauge symmetry group $SU(2)$. Herewith $U(1)$ plays the role of the invariant subgroup in $ SU(2)$; this, in turns, involves the nontrivial embedding of the holonomies group  $H$ "belonging" to the $U(1)$ gauge symmetry into the  holonomies group $H'$ "belonging" to the $SU(2)$ gauge symmetry. \par  
Those of  elements (\ref{cycl}) that 
correspond to  exact 1-forms at the just described embedding  form, in  turn,  the so-called \it
restricted holonomies subgroup \rm in the residual gauge symmetry group $U(1)$
\footnote{In some modern physical literature, e.g. in the papers \cite{Mitja1}, this group refers to as \it the trivial holonomies group\rm.}.   We  shall denote it as $\Phi^0$ henceforth. 
\par
It is easy to see that in the Minkowskian YM model \cite{LP1, Pervush2,David2,David1} the restricted holonomies  group $\Phi^0$  is isomorphic to the "small" subgroup $U_0$  in the residual gauge symmetry group $U(1)$ represented by Gribov topological multipliers $v^{(n)}({\bf x})$, (\ref{degeneration1}).\par
Indeed, only "small" Gribov  multipliers form the complete group $U_0$ ($\pi_0 (U_0)=0$; $\pi_1 (U_0)=0$), while each set of "large" Gribov  multipliers with the fixed topological number $n\in \bf Z$ is only a \it monoid\rm; the element inverse to the given one in the said "large" set with $n$ always belongs to the set with $-n$ ($n\neq 0$):
$$ v^{(-n)} ({\bf x})= [v^{(n)} ({\bf x})]^{-1}. $$  
One may  speak that the unit element of the holonomies group
$H$ is \it degenerated \rm  with respect to all the exact
forms corresponding to  zero topological charges.
In our case of  vacuum Higgs fields $\Phi_{(0)b}$ just
these fields  specify the class of exact forms and restricted holonomies  
group $\Phi^0$.\par
As it was shown in the monographs \cite{Al.S.},  when 
$A_\mu$ and $ A'_\mu$ are  two gauge fields the
gauge equivalence between which is realized through a 
function $g(x)\in G$, then
\be
\label{equiv}
b'_\gamma = g(x_1)b_\gamma g(x_1)^{-1}
\ee 
as the curve $\Gamma$ begins and ends in the point $x_1$
(\it the holonomies elements of  two gauge equivalent fields are
conjugate\rm).\par
Let $g(x)$ have the spatial asymptotic $g(x)\to 1$ as $x\to \infty$ (indeed,
$g(x)$ would have such asymptotic already on distances $\sim 1$ fm.; this is  associated
with the needs of the infrared quark confinement as we shall see below).
Let also $b_\gamma$ and $b'_\gamma$ be  elements of $H$
constructed by external forms belonging to a one class of
cohomologies. In conclusion, the gauge fields $A_\mu$ and $ A'_\mu$, 
forming the elements $b_\gamma$ and $b'_\gamma$ respectively (these fields "belong" to the $U(1)\to SU(2)$ embedding) and
connected by the gauge transformation $g(x)\to 1$,
\it are taken in the Coulomb  gauge \rm (\ref{transv}).\par
 Each such class
is obtained from the zero class
of exact forms \it as its Gribov copy \rm (thus we also "copy"    the
Gribov ambiguity equation (\ref{Gribov.eq}) written down in terms of Higgs fields with zero topological charges).
\par
As 
$g(x)\to 1$, $x\to \infty$, we may rewrite (\ref{equiv}) as
\be
\label{coh}
b'_\gamma =b_\gamma \cdot 1.
\ee
The latter equality reflects the structure of the
cohomologies group $H^1$: \it  two \rm 1-\it forms
belonging to one class of
cohomologies are equivalent to within an
exact form \rm (see \S T6 in \cite {Al.S.}). 
\footnote{We may directly make sure  that always there exists  the infinite set of transverse YM fields in each cohomological class. As two 1-forms $\omega_1\equiv A_\mu d^\mu$ and $\omega_1'\equiv A'_\mu d^\mu $,  involving the transverse YM fields $A_\mu$ and $ A'_\mu$ respectively, belonging to  the fixed cohomologies class differ on the exact 1-form $d\sigma$: $\omega_1-\omega'_1=d\sigma$, we may approximately write:
$$\partial_\mu (\omega_1-\omega'_1)= \partial_\mu d\sigma=0,$$
as far as $d\cdot d \sigma =0$ due do the Poincare lemma.\par
Moreover, one may be proved that always takes place  the actual \it duplication \rm of the infinite number of transverse YM fields in each cohomological class. It is the direct consequence of the Gribov ambiguity equation (\ref{Gribov.eq}) and nontrivial $U(1)\to SU(2)$ embedding. Applying the arguments similar to those in the work \cite{Baal}, we may always choose two infinite close to each  other transverse YM fields (in each topological sector of the Minkowskian YM model \cite{LP1, Pervush2,David2,David1}) in such a wise that one of these fields belongs to the Lee algebra of $SU(2)$, while another one belongs to the Lee algebra of $U(1)$.}
In terms
of the holonomies group $H$ this, in turn,  implies that
 two elements of $H$ corresponding to the 1-forms
belonging to a one class of
cohomologies are \it equivalent  to within an 
element of the restricted holonomies group $\Phi^0$ \rm. \par
Thus the Gribov ambiguity equation (\ref{Gribov.eq}) and  Gribov gauge
transformations (\ref{degeneration})
correctly describe  the nontrivial
chorological structure of the transverse vacuum YM fields satisfying the 
Coulomb  gauge \rm (\ref{transv}) 
 and involving the 
 non-Abelian gauge
group $SU(2)$ then broken down spontaneously  to
the $U(1)$ gauge
group. 
\par
Note that specifying the above  cohomological structure of transverse YM fields,
we herewith
solve, in fact, the Cauchy task (\ref{Gribov.eq}), (\ref{Gauss2}),  (\ref{Aparallel}) (\it expanded on all the cohomologies classes due to the Gribov topological mechanism \rm (\ref{degeneration}) \it and Pontryagin theory \rm (\ref{deg})), i.e. define, in the given initial time instant $t_0$, the space-like surface  ${\cal H}(t_0)$ in the Minkowski space-time where the Gribov topological degeneration of initial  data occurs.\par
One may  show  \cite {David2,David1} that  the Gribov equation (\ref{Gribov.eq}) together with
the topological condition
\be
\label{X[n]}
X[\Phi ^{(n)}]=n \ee
are compatible with the unique solution to the classical equations.
\it They just result   Wu-Yang monopoles $\Phi ^{(n)}_k$ considered above\rm. \par
The nontrivial solution to the equation for the Gribov phase in this case is well known \cite {LP1,Pervush2}:
\be
\label{phase}
{\hat \Phi}_0(r)= -i\pi \frac {\tau ^a x_a}{r}f_{01}^{BPS}(r),
\quad f_{01}^{BPS}(r)=[\frac{1}{\tanh (r/\epsilon)}-\frac{\epsilon}{r} ].
 \ee
It is just a $U(1)\to SU(2)$ isoscalar made of  Higgs vacuum BPS monopoles (\ref{sc monopol}).   \par
As a definite linear combination of Higgs vacuum BPS monopole modes 
(\ref{sc monopol}), the Gribov phase ${\hat \Phi}_0$ satisfies the Gribov ambiguity equation (\ref{Gribov.eq}) \cite {LP1, Pervush2}.
\par
 As a result, instead of  the topological degenerated classical vacuum inherent in the
instanton theory \cite {Bel} (in the physically unattainable region),
we get   a topologically degenerated Wu-Yang (BPS) monopole\rm:
\be
\label{mon.deg}
{\hat \Phi_i} ^{(n)}:= v^{(n)}({\bf x})[{\hat \Phi_i} ^{(0)}+\partial _i]v^{(n)}({\bf x})^{-1},\quad
v^{(n)}({\bf x})=
\exp [n\hat \Phi _0({\bf x})], \ee
and  topological degenerated multipoles:
\be
\label{mult}
{\hat {\bar A}}^{(n)}:=v^{(n)}({\bf x}){\hat {\bar A}}^{(0)}v^{(n)}({\bf x})^{-1}.
\ee
As regards Higgs vacuum BPS monopoles, they are also topologically degenerated and herewith in the same wise that YM multipoles \cite{Arsen}:  
\be \label{Higgsi}
\Phi_{(n)b}= v^{(n)}({\bf x}) \Phi_{(0)b} v^{(n)}({\bf x})^{-1}.
\ee
Let us now examine the behaviour of  the Gribov phase (\ref{phase}) at the origin of coordinates and at the spatial infinity. \par
The BPS ansatz  $ f_{01}^{BPS}(r) =f_0^{BPS}(r)~\epsilon$ in this formula becomes 1 as $r\to \infty$  \cite{Pervush2,David1}, while at the origin of coordinates it goes to zero.  \par 
It turns out  that it is a sign of a good physics. \par
Really, in this case
Gribov exponential multipliers
$$ v^{(n)}({\bf x})=\exp [n\hat \Phi _0({\bf x})]$$ 
become 1 at the origin of coordinates. \par 
On the other hand, at the spatial infinity the
 Gribov factors $ v^{(n)}({\bf x})$ would have the same asymptotic, to ensure the topological and quark confinements in the infrared region of  the momentum space \cite{LP1,Pervush2, David2, Pervush3, Ilieva, Nguyen,Azimov} via the complete destructive interference of  "large" Gribov multipliers $ v^{(n)}({\bf x})$ ($n\neq 0$) at the spatial infinity. \par
The features of the infrared topological (and quark) confinement \cite{Pervush2, David2, Pervush3, Ilieva, Nguyen, Azimov} will be us discussed below. But now we would like to consider the conditions that is necessary to impose onto ("large") Gribov multipliers $ v^{(n)}({\bf x})$ to achieve their  spatial asymptotic 
\be \label{bondari} v^{(n)}({\bf x})\to 1, \quad {\rm as} ~~\vert {\bf x}\vert \to \infty,\ee
providing the infrared topological  confinement. \par
Let us slightly modify  Eq. (\ref{phase}) for the Gribov phase 
in comparison with that in the works \cite{LP1, Pervush2,David1}. The paper  \cite{Azimov} prompts us how to do this. \par 
So, we rewrite Gribov exponential multipliers $ v^{(n)}({\bf x})$ as \cite{Azimov} 
\be \label{newfact}
v^{(n)}({\bf x})= \exp(\hat \lambda~_ {n,\phi_i} ({\bf x})), \ee 
where
\be \label{ln}
\hat\lambda~_{n,\phi_i} ({\bf x})\equiv 2i \tau ^a \Omega _{ab}(\phi_i)  \frac{x^b}{r} f_{01}^{BPS}(r)~ \pi n
\ee 
and 
$$ (\tau ^a)^\alpha_\beta ~\Omega _{ab}(\phi_i)= (u(\phi_i))_\gamma ^\alpha 
(\tau ^a)^\gamma _\delta (u^{-1}(\phi_i))_\beta ^\delta, $$  
\be \label{eiler} (u(\phi_i)) ^\alpha_\beta = (e^{i\tau_1 \phi_1 /2})_\gamma ^\alpha  (e^{i\tau_2 \phi_2 /2})^\gamma _\delta (e^{i\tau_3 \phi_3 /2})_\beta ^\delta.
\ee 
Here $\phi_i$ ($i=1,2,3$) are three Euler angles, fixing the position of the 
coordinate system in the $U(1)\to SU(2)$ group space.  \par
To obtain the necessary asymptotic (\ref{bondari}) at the spatial infinity, we should impose the appropriate conditions onto the  Gribov phase in the modified Eq. (\ref{newfact}). More precisely, such conditions may be imposed onto Euler angles $\phi_i$. \par
\it Gribov topological factors $v^{(n)}({\bf x})$ become \rm 1 \it at the spatial infinity as \rm
\be \label{fas.ysl-e} \tau_i \phi_i = 4\pi n, \quad n\in {\bf Z}.
\ee
\subsection{ Topological dynamics and chromo-electric monopole.}  
 Gribov copies of YM fields belonging to the zero topological sector of the Minkowskian theory involving  vacuum BPS monopoles are an evidence of the zero mode in the left-hand side of
the  both  constraints (\ref{Gauss}), (\ref{Gauss1}):
\be
\label{zero mode}
[D^2_i(\Phi ^{(0)})]^{ac} A_{0c}=0.
\ee
The nontrivial solution to this equation,
\be
\label{sol.zero}
A_0^c(t,{\bf x})= {\dot N}(t)\Phi_0^c ({\bf x})
\ee
(it is, indeed, the solution of the Gauss law constraint (\ref{Gauss}) additional \cite{Pervush1}  to the trivial one, $A_0=0$, obtained  at
the Dirac removal \cite{Dir} of temporal YM components),
may be removed from the local equations of motion by the gauge transformation (\ref{udalenie}) (a la Dirac in 1927 y. \cite{Dir}) to convert
YM   fields \it into topological
Dirac variables\rm:
\be
\label{d.v.}
\hat A_i^{(N)}=\exp [N(t)\hat \Phi_0 ({\bf x})]
[\hat A^{(0)}_i+\partial _i]\exp [-N(t)\hat \Phi_0 ({\bf x})].
\ee
But the solution (\ref{sol.zero}) cannot be removed from the constraint-shell action 
$$W^*= \int dt ~{\dot N}^2(t) I/2+\dots$$ 
and from the winding number functional 
$$X[A^{(N)}]=N(t)+X[A^{(0)}]$$ 
(the
latter equality implies that local YM degrees of freedom $A^{(N)}$ are completely separated  
from $N(t)$ \cite{LP1}).
\par
Finally, we get the Feynman path integral
\be
\label{path i} Z_F=\int DN \prod \sb {i,c}  [D E^{c(0)}_{i} D A^{i(0)}_{c}]e^{iW^*},
\ee
that contains the additional topological variable $ N(t)$.\par
We shall consider  deriving  the integral (\ref{path i}) in the next sections.\par
In the lowest order of the 
perturbation theory the constraint (\ref{zero mode})  permits the solution (\ref{sol.zero}), implying a
vacuum \bf "electric"  monopole \rm 
\be \label{el.m}
F^b_{i0}={\dot N}(t)D ^{bc}_i(\Phi_k ^{(0)})\Phi_{(0)c}({\bf x}).
\ee
We call the just appeared variable $N(t)$ \it the  winding number variable\rm.
 It is specified with the aid of the  vacuum Chern-Simons functional (in the zero topological sector),
 equal to the difference of the \it in \rm and  \it out \rm
values of this variable:
\bea
\label{winding num.}
\nu[A_0,\Phi^{(0)}]&=&\frac{g^2}{16\pi^2}\int\limits_{t_{in} }^{t_{out} }dt
 \int d^3x F^a_{\mu\nu} \widetilde{F}^{a\mu \nu}=\frac{\alpha_s}{2\pi}
 \int d^3x F^b_{i0}B_i^b(\Phi^{(0)})[N(t_{out}) -N(t_{in})]\nonumber \\
 &&
 =N(t_{out}) -N(t_{in}).
\eea 
The winding number functional admits its generalization \it to 
noninteger degrees of the map \rm \cite {Niemi}:
\be
\label{Xni}
X[\Phi^{(N)}]\neq n ~(n \in {\bf Z}),\quad  ({\hat \Phi}^{(N)}=
e^{N{\hat \Phi}_0}[\hat \Phi_i ^{(0)}+\partial _i]e^{-N{\hat \Phi}_0}).
\ee
We may identify the global variable $N(t)$ with the winding number 
degree of freedom in the Minkowski space associated with the  \it free rotator action \rm
\be
\label{rot}
W_N=\int d^4x \frac {1}{2}(F_{0i}^c)^2 =\int dt\frac {{\dot N}^2 I}{2},
\ee
 involving the rotary momentum  (cf. (4.6) in  \cite{David2}):
\be
\label{I}
I=\int \sb {V} d^3x (D_i^{ac}(\Phi_k)\Phi_{(0)c})^2 = \frac {4\pi^2\epsilon}{\alpha _s}
=\frac {4\pi^2}{\alpha _s^2}\frac {1}{ V<B^2>},
\ee
 does not
contribute  to the local equations of motion. This
free rotator action disappears in the limit $V\to \infty$ ($\epsilon \to 0$). \par
The influence of the infinite mass of the
Higgs field, $m/\sqrt{\lambda}$, onto the free rotator action (\ref{rot}) reduces, at the spatial infinity 
($V\to \infty$),  in fact to zero, as we learn 
this from (\ref{I}). \par
The rotation action (\ref{rot}) describes the collective solid potential rotation of the Minkowskian 
(YM-Higgs) vacuum \cite{LP1, Pervush2,David2, David1} \rm (the latter one is, indeed, a kind of superfluid Bose condensate \cite{N.N.}, alike to the superfluid component in a helium II specimen). 
\par
Our reasoning about $<B^2>\neq 0$ (this value is induced by the Bogomol'nyi  equation (\ref{Bog}) and characterizes the superfluid 
(YM-Higgs) vacuum), 
confirms that the action (\ref{rot}), depending on the \it collective topological variable \rm $N(t)$ via its temporal derivative $\dot N(t)$ and describing the  collective solid potential rotation of the Minkowskian 
(YM-Higgs) vacuum \cite{LP1, Pervush2,David2, David1} (in fact,  the action (\ref{rot}) is induced by vacuum Higgs BPS monopole modes (\ref{sc monopol})),
is a purely Minkowskian 
 effect\rm.\par 
Now,  taking account of  evaluating  the magnetic energy
(\ref{magn.e}), we may write down the action of the of the Minkowskian YM theory \cite{LP1, Pervush2,David2, David1} in the lowest order of the 
perturbations theory, describing only the (physical) vacuum of this  theory. \par
This  
action \cite{LP1}  contains the both kinds of vacuum BPS monopoles,  the "electric" and the "magnetic" ones: 
\be
\label{ca}
W_{\cal Z} [N,\Phi^{0 BPS}]= \int dt d^3x \frac {1}{2}\{ [F^b_{i0}]^2-
[B^b_i(\Phi^{0BPS})]^2\}= \int dt\frac {1}{2}\{ I\dot N ^2- 
\frac {4\pi}{g^2 \epsilon}\}.
\ee 
The topological degeneration
of all the
fields reduces to the degeneration of the global topological
variable $N(t)$ with respect to the shift of this variable on integers:
$$N \Longrightarrow N+n ;\quad n= \pm 1,\pm 2,\dots;\quad 0 \le N(t)\le 1.$$
Thus the topological variable $N(t)$
 specifies the  free rotator  (\ref{rot}) involving  the instanton-type wave function (\ref{plan})
of the topological motion in
the Minkowski space-time:
\be
\label{PsiN}
\Psi _N=\exp (iP_N N), \quad P_N ={\dot N} I= 2\pi k +\theta,
\ee 
with $k$ being  the \it number of the Brillouin zone \rm
and $\theta$ being the $\theta$-angle (or \it  the   Bloch quasi-momentum\rm)
\cite{LP1,Pervush1,Pervush2}. \par 
The action (\ref{ca}) of the Minkowskian YM theory \cite{LP1, Pervush2,David2, David1}  in the lowest order of the perturbations theory induces
the  appropriate vacuum Hamiltonian (written down in terms of the canonical 
momentum $ P_N =\dot N I$) \cite{LP1}: 
\be
\label{Hamilton}
H= \frac {2\pi}{g^2\epsilon}[ P_N^2 (\frac {g^2}{8\pi^2})^2+1].
\ee  
We see that this Hamiltonian, depending on $ P_N^2$, is explicitly \it Poincare invariant \rm (in particular, CP invariant), unlike the $\theta$-term in the effective  Lagrangian  of
the instanton YM theory \cite{Bel}. \par
Thus the well-known CP-problem \cite{Cheng} may be  solved in the Minkowskian YM  theory  \cite{LP1, Pervush2,David2, David1}  involving vacuum 
"electric" monopoles (\ref{el.m})\rm. \par
In contrast to the instanton wave
function (\ref{plan}), inherent in the Euclidian YM theory \cite{Bel} and involving \it the purely imaginary energy-momentum spectrum \rm \cite{LP1, Pervush2,Arsen,Galperin},
the spectrum of the topological momentum $P_N$ in  the Minkowskian
YM theory \cite{LP1, Pervush2,David2, David1} \it is  real  and belongs to
the  physical values\rm. \par
Finally, Eqs. (\ref{rot}), (\ref{PsiN}) specify \it the  countable spectrum \rm of the global "electric" tension (\ref{el.m}):
\be
\label{el.m1}
F^a_{i0} =\dot N(t) ~(D_i (\Phi_k^0) \Phi_0)^a= P_N \frac {\alpha_s}{\pi^2\epsilon} B_i^a (\Phi _0)=\vert 2\pi k +\theta\vert \frac {\alpha_s}{\pi^2\epsilon} B_i^a(\Phi_0).
\ee
It is an analogue of the Coleman spectrum of the electric tension in the
$QED_{(1+1)}$ \cite {Coleman}:
\be
\label{Col.sp}
G_{10}=\dot N \frac {2\pi}{e} =e(\frac {\theta}{2\pi}+k).
\ee
The application of the Dirac quantization to one-dimensional
electrodynamics $QED_{(1+1)}$ in the paper \cite{Gogilidze}
demonstrates the universality of (topological)  Dirac variables and their
adequacy to the description of  topological dynamics in terms of
a nontrivial homotopies group.
\section {Zero mode of Gauss law constraint.}
\subsection {Dirac variables and zero mode of Gauss law constraint.}
The Minkowskian YM constraint-shell theory \cite{LP1, Pervush2,David2, David1}   is obtained at  explicit resolving  the
Gauss law constraint (\ref{Gauss}), and our  next step is associated  with the
initial action on the surface of these solutions:
\be
\label{in.act}
W^*= W[A\mu] \vert _ {\frac {\delta W}{\delta A_0}=0}.
\ee
The results of  similar  resolving in QED are \it  electrostatic and
 Coulomb-like atoms\rm. \par
In the non-Abelian case the
topological degeneration in the  form of Gribov copies (\ref{degeneration1})
 implies that
the  general solution to the Gauss law constraint  (\ref{Gauss}) contains the zero
mode ${\cal Z}$ as a peculiar indicator of the  Coulomb gauge (\ref{Aparallel}).\par   
 The general solution to  the   heterogeneous equation (\ref{Gauss}) may be represented \cite{LP1, Pervush1} as
the sum of the zero-mode solution ${\cal Z}^a$
to the homogeneous equation
\be
\label{hom}
(D^2 (A))^{ab}{\cal Z}_b =0 
\ee
and a particular solution, ${\tilde A}_0^a$, to the heterogeneous one, that is
\be
\label{inhom}
A_0^a={\cal Z}^a+{\tilde A}_0^a.
\ee
On the other hand,  the zero-mode ${\cal Z}^a$, may be represented, at  the spatial infinity, 
 in the form of the sum of the product of the new topological variable
${\dot N}(t)$  onto the Gribov phase $\hat\Phi _{(0)}({\bf x})$, (\ref{phase}), and weak
multipole corrections:
\be
\label{Summ}
{\hat {\cal Z}}(t,{\bf x})\vert _{\rm asymptotic}= {\dot N}(t)\hat\Phi _{(0)}({\bf x})+O (\frac {1}{r^{l+1}}),
\quad (l>1).
\ee In this case the single one-parametric variable $N(t)$ reproduces the
topological degeneration of all the YM field  as the topological Dirac variables
are specified by the gauge transformations
\be
\label{tr.d}
0= U_{\cal Z}({\hat {\cal Z}}+ \partial _0)U_{\cal Z}^{-1}
\ee
\be
\label{tr.d1}
{\hat A}^*_i =U_{\cal Z}({\hat A}_i^0+\partial _i)U_{\cal Z}^{-1}, \quad 
{\hat A}_i^{(0)} =\Phi _i^{(0)}+{\bar A}_i^{(0)},
\ee
where the spatial asymptotic of $U_{\cal Z}$ is 
\be
\label{spat.as}
U_{\cal Z}= T \exp [ \int ^t dt'{\hat {\cal Z}}(t',{\bf x})]\vert _{\rm asymptotic}=
\exp [N(t)\hat\Phi _{(0)}({\bf x})].
\ee
The topological degeneration of all  the fields reduces in this case to the
degeneration of only one global topological variable $N(t)$ with
respect to the
shift of this variable on integers: $(N \Longrightarrow N+n,n= \pm 1,\pm 2,...)$.
\subsection { Constraining with  zero mode.}
Let us  specify \it the equivalent unconstrained system \rm for the Minkowskian YM theory \cite{LP1, Pervush2,David2, David1}   in
the monopole class of functions in the presence of the zero mode
${\cal Z}^b$ of the Gauss law constraint (\ref{Gauss}), (\ref{hom}):
\be
\label{A}
A_0^a={\cal Z}^a+{\tilde A}_0^a ;\quad F^a_{0k}=
- D^{ab}_k(A){\cal Z}_b+{\tilde F}^a_{0k} \quad (( D^2(A)^{ab}{\cal Z}_b =0 ).
\ee
To get the constraint-shell action:
\be
\label{constr}
W_{YM}({ constraint}) = {\cal W}_{YM}[{\cal Z}]+{\tilde W}_{YM}[{\tilde F}],
\ee
we use the obvious decomposition:
\be
\label{relations} F^2=(-D{\cal Z}+{\tilde F})^2= (D{\cal Z})^2- 2{\tilde F}D{\cal Z}+{\tilde F}^2= 
\partial ({\cal Z}\cdot D{\cal Z})-
2 \partial ({\cal Z}{\tilde F})+({\tilde F})^2.
\ee
The latter relation is true due to the Bianchi identity $D {\tilde F}=0$,
the Gauss law constraint  equation $D ^2{\cal Z}=0$ and the explicit expression for the
derivative $D{\cal Z}$:
$$D{\cal Z}= (\partial {\cal Z}+ g {\bf A}\times{\bf {\cal Z}}).$$
This shows that the zero mode part, ${\cal W}_{YM}[{\cal Z}]$, of
the constraint-shell action (\ref{constr}) is, indeed, the sum of  two integrals:
\be
\label{int.sum}
{\cal W}_{YM}[{\cal Z}]= \int dt \int d^3x [\frac {1}{2} \partial_i ({\cal Z}^a D^i_{ab}(A){\cal Z}^b)-
\partial_i (\tilde F^a_{0i}
{\cal Z}_a)] ={\cal W}^0+{\cal W} ', \ee
where  the first term,${\cal W}~^0$, is  the \it  action of a free rotator\rm,
(\ref{rot}),
 and the
second one, ${\cal W} ~'$,
describes \it the coupling of the zero-mode to  local excitations\rm. \par
These  terms are  specified by the asymptotic of  fields
$({\cal Z}^a,~A^a_i)$ at the spatial infinity. We shall denote them as ${\dot N}(t)\Phi _{(0)}^a({\bf x})$ and $\Phi _i^a({\bf x})$ respectively.
The fluctuations $\tilde F^a_{0i}$ belong  to the class of multipoles. \par
 As long as the surface integral over  the monopole-multipole couplings vanishes
(due to the  Gauss-Ostrogradsky theorem and  spatial asymptotic (\ref{Summ}) for the zero mode solution), the fluctuation part
 of the second term  drops out: there are no  contributions from the interference between the "electric" monopole and perturbation excitations: multipoles and Higgs "electric" excitations, in the constraint-shell action (\ref{constr})\rm. \par
Thus \cite {LP1}  scalar BPS monopoles disappear, in effect, from the excitations spectrum in the 
infinite spatial volume limit \rm $V\to\infty$. \par
Substituting 
the solution with
the asymptotic (\ref{Summ}) into  the first  term of  Eq. (\ref{int.sum}) implies  the
free rotator  action (\ref{rot}).\par The action for the equivalent unconstrained system of local excitations
 (cf. Eq. (21) in \cite {Pervush2}):
\be
\label{unconstr}
\tilde W_{YM}[\tilde F]= \int d^4x\{ E_k^a \dot A_a^{k(0)}- 
\frac {1}{2} \{E_k^2+B_k^2 (A^{(0)})+[D_k^{ab}(\Phi^{(0)})\tilde \sigma_b]^2\}\},
\ee
is obtained in terms of  variables with the zero degree of the  map:
\be
\label{zero}
\hat{\tilde F}_{0 k}=U_{\cal Z}{\tilde F}_{0 k}^{(0)}U_{\cal Z}^{-1},\quad 
\hat{A}_i=U_{\cal Z}(\hat{A}_i^{(0)}+\partial_i)U_{\cal Z}^{-1},
\quad  \hat{A}_i^{(0)}(t,{\bf x})=\Phi_i^{(0)}(t,{\bf x})+\hat{\tilde A}^{(0)}(t,{\bf x}),
\ee
by  decomposing   the "electric" component of the  YM  tension
tensor $F _{0 i}^{(0)}$ into their  transverse: $E_i^a$, and longitudinal:
$$F _{0 i}^{aL}=- D_i^{ab}(\Phi ^{(0)})\tilde \sigma_b,$$ 
parts; so 
\be
\label{F0}
F _{0 i}^{a(0)}=E_i^a-D_i^{ab}(\Phi ^{(0)})\tilde \sigma_b.
\ee
Here the function $\tilde \sigma^b$ is specified by the Gauss law equation (\ref{Gauss}) rewritten in terms of Dirac variables:
\be
\label{G.e}
((D^2(\Phi ^{(0)}))^{ab}+ g\epsilon ^{adc}\tilde A _{id}^{(0)}D^{ib}_c(\Phi ^{(0)}))\tilde \sigma_b =- g\epsilon ^{abc}
\tilde A _{ib}^{(0)}E^i_c
\ee
(we  recommend our readers the monograph \cite{Slavnov}, p. 88,
where Eqs. (\ref{F0}), (\ref{G.e}) were derived in the Hamiltonian formalism of the YM
theory). \par
The  transverse "electric"
tension
tensor $E_i^a$ appears in the considered Minkowskian YM theory.  We may write   down it  in terms of  the zero mode ${\cal Z}^a$ of the Gauss law constraint (\ref{hom}) \cite{David2}:
\be \label{pop.napr}
E_i^a= D_i^{ac} (A) A_{0ac},
\ee
with $A_0^a$  given in  (\ref{inhom}).  Due to (\ref{inhom}), the "electric" tension
tensor $E_i^a$
may  be, in turn, decomposed into the zero  mode part and  perturbations one,
${\tilde E}_j^c$:
\be \label{edg} E_i^a= D_i^{ac} {\cal Z}_c + {\tilde E}_i^a;
\ee 
the  latter item enters  the unconstrained excitation action (\ref{unconstr}).\par 
In order to construct the consequent QCD in the Minkowski space, we should incorporate fermions (quarks) in  that theory. We also consider them  as  perturbation excitations over the Minkowskian physical (YM, gluonic) vacuum \cite{LP1, Pervush2,David2, David1}. (we  shall treat the latter one as the ground level of the perturbations theory). Thus we may 
supplement the unconstrained excitation action (\ref{unconstr})
by the fermionic contribution \cite{David2, David1}: 
\be
\label{fermioni}
\tilde W [\Psi] = \int d^4 x \bar\psi[i\gamma^\mu(\partial _\mu+{\hat A_\mu})
-m]\psi. \ee
When we then introduce the current $j$ of independent non-Abelian
variables \cite{LP1}: \be
\label{cur1}
j_0 ^a = g \epsilon ^{abc} [A^{*a(n)}_{ib}-\Phi_{ib}^{a(n)}]\tilde E ^{i}_c, 
\ee
the  fermionic  term  (\ref{fermioni}) allows us  to consider the so-called \it total current\rm: the sum of the 
current $j$,  (\ref{cur1}), of independent non-Abelian
variables, and a fermionic  current \cite{David1}: 
\be \label{total current}
j^b_{tot~0}=g\epsilon^{abc}{\tilde A}_i^{a} \tilde E_i^{c}+j^{fb}_0,
\ee
with $j^{fb}_0$ being a fermionic current. \par
Then 
Eq. (\ref{G.e}) may be rewritten as
\be
\label{cur-t}
D_i^{cd}(A)D^i_{db}(\Phi^{(0)})\tilde \sigma ^b = j_0 ^c.
\ee
The latter equation is wholly determined  by the  perturbation excitations over the  zero mode $\hat{\cal  Z}$, (\ref{Summ}).
\par
Due to  its manifest gauge invariance, the dependence of the action for local
excitations on the zero mode
disappears, and we get the ordinary generalization of the
 covariant  Coulomb gauge \cite {Schwinger,Fadd1,Gitman}
 in the presence of vacuum  Wu-Yang (BPS) monopoles.
\section { Rising potential induced by Wu-Yang monopoles.}
Now we  are able to calculate the Green function of the Gauss constraint equation (\ref{cur-t}),
 coinciding mathematically with the Gribov   equation (\ref{Gribov.eq})
 (see \cite {David2}, \S 4.C):
\be
\label{Gr.eq}
D^2((\Phi ^{(0)})^{ab}({\bf x})G_b^c ({\bf x},{\bf y})=
\delta^{ac}\delta^3(x-y),
\ee
that sets, due to (\ref{total current}),  (\ref{cur-t}), the  potential of the current-current instantaneous interaction:
\be
\label{cint}
-\frac {1}{2} \int \sb{V_0} d^3 x d^3 y j_{tot,0} ^b ({\bf x})
G_{bc} ({\bf x},{\bf y})j_{tot,0}^c({\bf y}).
\ee
Note  \cite {David1} that Eq. (\ref{Gr.eq}) allows us to find the "longitudinal" function $\tilde \sigma _b$
as an expansion by powers of $g$:
\be \label{fse} \tilde \sigma^{b}(t,{\bf x})=
- \int d^3yG^{bc}({\bf x},{\bf y})j_{tot,0}^c(t,{\bf y}) -
\int d^3yd^3zG^{bc}({\bf x},{\bf y})gf^{cde}{\tilde A}_k^{d}(t,{\bf y})
G^{ef}({\bf y},{\bf z})j_{tot,0}^{f}(t,{\bf z}) - \dots ,
\ee 
by analogy with the standard perturbations theory \cite{Abers}.
\par
In the presence of a vacuum  Wu-Yang monopole we  get 
\be
\label{Gr.eq.mon}
D^2((\Phi ^{(0)})^{ab}({\bf x})
= \delta^{ab}\Delta -\frac {n^a n^b+\delta^{ab}}{r^2}+2(\frac {n^a}{r}\partial ^b-\frac {n^b}{r}\partial ^a),
\ee
with 
$$n_a(x)=x_a/r; \quad r=\vert {\bf x}\vert.$$ 
Let us decompose $G^{ab}$  into the complete set of  orthogonal
vectors in the colour space:
\be
\label{complete set}
G^{ab}({\bf x},{\bf y})= [n^a(x) n^b(y)V_0(z)+ \sum \sb {\alpha=1,2} 
e^a_ \alpha (x)e^{b\alpha}(y)V_1(z)];\quad (z=\vert {\bf x}-{\bf y }\vert).
\ee
Substituting the latter  expression into Eq.  (\ref{Gr.eq}),  we get
\it the Euler equation \rm (see \cite {Kamke}, Eq. (2.160)):
\be
\label{Euler}
\frac {d^2}{dz} V_n+ \frac {2}{z}\frac {d}{dz}V_n- \frac {n}{z^2}V_n =0, \quad n=0,1.
\ee
The general solution to the latter equation is
\be
\label{V}
V_n (\vert {\bf x}-{\bf y} \vert)=d_n\vert {\bf x}-{\bf y} \vert ^{l^n_1}+c_n\vert {\bf x}
-{\bf y} \vert^{l^n_2}, \quad n=0,1 ,
\ee
with $d_n,~c_n$ being constants, while $l^n_1,~l^n_2$
may be found as the roots of the equation $$l^{n2}+l^n=n,$$
 i.e.
\be
\label{roots}
l^n_1= -\frac {1+\sqrt{1+4n}}{2};~~~~~\quad l^n_2=\frac {-1+\sqrt{1+4n}}{2}.
\ee
It is easy to see that for $n = 0$, as $d_0=-1/4\pi$, we get
 the \it Coulomb-type potential\rm:
\be
\label{Coulomb}
l^0_1= -\frac {1+\sqrt{1}}{2}=-1 ;\quad l^0_2=\frac {-1+\sqrt{1}}{2}=0,
\ee
\be
\label{Cp}
V_0 (\vert {\bf x}- {\bf y} \vert) = 
-1/ 4\pi ~\vert {\bf x}- {\bf y} \vert ^{-1} + c_0;
\ee
and for $n = 1$, the "\it golden section\rm" potential for the 
\it golden-section equation \rm 
$$(l^1)^2+l^1=1,$$ 
with
\be
\label{ris}
l^1_1= -\frac {1+\sqrt{5}}{2}\approx -1.618;\quad l^1_2=\frac {-1+\sqrt{5}}{2}\approx 0.618,
\ee
\be
\label{ris1}
V_1 (\vert {\bf x}-{\bf y} \vert)=
-d_1\vert {\bf x}-{\bf y} \vert ^{-1.618}+c_1\vert {\bf x}-{\bf y} \vert^{0.618}.
\ee
The latter, "golden section", potential (unlike  the Coulomb-type one)
implies the rearrangement of the naive perturbation series and the spontaneous breakdown of the
chiral symmetry. In turn, this involves the  constituent gluonic mass  in the Feynman graphs: this mass changes
the asymptotic freedom  formula in the region 
of  low transferred  momenta; thus the
coupling constant $\alpha _{QCD}(q^2\sim 0)$ becomes finite.  \par
The  "golden section" potential (\ref{ris1})
 may be also considered
 as an origin of  "\it hadronization\rm" of  quarks and
gluons in QCD \cite {Pervush2,David2,Bogolubskaja,Werner,Yura2}.\par
Our assumption  that the  "golden section" potential (\ref{ris1}) changes the infrared behaviour of Minkowskian QCD and gluondynamics has its roots in one very illustrative and herewith simpler example of  increasing potentials.  \par
Let us consider, following the papers \cite {Bogolubskaja,Werner}, the  squared potential \cite{Pervush2,Yao}
\be \label{Ya}
V_R(r)= V_0 r^2; \quad V_0^{1/3} \sim  234 {\rm MeV}.
\ee
In  the papers \cite {Bogolubskaja,Werner} there was demonstrated, by 
application of definite numerical methods (there were the  "shooting" \cite{6} and  the
Runge-Kutta-Gill   \cite{7} methods)   that the appearance of the squared increasing potential (\ref{Ya}) \cite{Yao} involves the following 
Modification of the gluonic propagator in the infrared momenta region \cite {Bogolubskaja,Werner}:
\be  \label{modif} D_{ij}^{mod} (k_0,k)= (\delta_{ij} - \hat k_i\hat k_j )\frac{1}{k_0^2- \omega^2 ({\bf k}) -i\epsilon},
\quad   \hat k_i =  \frac{1}{\vert {\bf k}\vert} k_i: \ee 
with 
$$ 
\underline \omega (k) \longrightarrow \frac{2}{\underline{\bf k}^2}, \quad {\bf k}\to 0,
$$
 \be 
\label{modiff}  \underline \omega (k) \longrightarrow \underline{\bf k}\quad  {\bf k}\to \infty; \ee
where 
\begin{eqnarray}\label{modiff1}  
 \left( \begin{array}{ll} \underline \omega\\ \underline k   \end{array} \right) =
(N_cV_0)^{-1/3}  \left( \begin{array}{ll}  \omega\\  k   \end{array} \right) 
 \end{eqnarray}
and $N_c$ is the number of colours in the considered theory.   This implies
that gluons effectively  acquire a structural mass depending on the
momentum: 
\be \label{constmass}   m_g(k ^2)= \sqrt{\omega^2(k) -{\bf k}^2}, \ee
such that $m(0)\to \infty$. \par
The modified gluonic propagator (\ref{modif}) vanishes in this limit. \par 
On the other  
hand, in the ultra-relativistic  region ${\bf k}\to \infty$ we come to the standard gluonic propagator providing the asymptotic freedom of quarks. \par
It is also worth  to cite here \cite{Bogolubskaja, Werner}  the "modified" formula for the running gluonic coupling constant $\alpha_s^{mod} (q)$ (as a function of the transferred momentum $q$) that follows from the formula (\ref{modif}) for the "modified" gluonic propagator in the presence of the square increasing potential  $V_R(r)$ \cite{Pervush2,Yao}.\par
In the "world without quarks" i.e. when the number of flavours $N_f=0$, the "modified" formula for the running gluonic coupling constant $\alpha_s^{mod} (q)$ takes the look \cite{Werner}
\be \label{almod} {\alpha}_s^{mod}(0) =
\frac{1}
{
\beta  \left[
1  +  \ln \left( \displaystyle {
\frac{ 4N_c  V_0^{1/3}} {\Lambda} }  \right)^2
\right] }
\approx  0.2;    \quad \beta     =     \frac{11}{4 \pi};\ee
at the zero value of the transferred momentum $q$. Herewith the ultraviolet 
cut-off parameter $\Lambda$ is adopted \cite{Bogolubskaja} to be  $\Lambda \approx 110$ Mev. \par
There may be shown \cite{Bogolubskaja, Werner} that $\alpha_s^{mod} (q)\leq  \alpha _{s}^{mod}  \left( 0 \right)  \stackrel{<}{\sim}  0.2  $ in the whole permissible region of the transferred momentum $q$ and,  therefore,  one  may  use
the  perturbation  theory  for  all the  transferred  momenta  $q$.  \par
The precise computations about  the   gluonic propagator and gluonic coupling constant in the Minkowskian non-Abelian  theory \cite{LP1, Pervush2,David2, David1}  also demand the application of definite numerical methods,  alike the methods \cite{6,7} that were applied \cite{Bogolubskaja, Werner}  in the case of the squared potential (\ref{Ya}) \cite{Yao}. We leave these  computations out of our present discussion.  
\section {Feynman and FP path integrals.}
In the "pure" Minkowskian YM theory \cite{LP1, Pervush2,David2, David1}, without of the fermionic sector, the Feynman path integral over  independent variables includes, in particular, the  integration
over the topological variable $N(t)$:
\be
\label{N.i}
Z_F[J]= \int \prod \sb {t} dN(t)\tilde Z[J^U],
\ee
with
\be
\label{Z}
\tilde Z[J^U]=\int \prod \sb {t,x} \{\prod \sb {a=1} ^3 \frac {[d^2 A_a^{(0)}d^2 E_a^{(0)}]}{2\pi} \}
\exp i\{{\cal W}_{YM}({\cal Z})+\tilde W_{YM}(A_a^{(0)})+S[J^U]\}.
\ee
As we have seen above, the functionals $\tilde W$ and $S$
are given in terms of  variables containing  non-perturbation
phase factors $U=U_{\cal Z}$, (\ref{spat.as}),
of the Gribov topological degeneration of  initial data.
These factors disappear in the action functionals $W_{YM}$ and $\tilde W_{YM}$ (due to their absorption \cite{Cheng} in the gauge invariant YM tensor squared, $(F_{\mu\nu}^a)^2$, on which the action functionals $W_{YM}$ and $\tilde W_{YM}$ explicitly depend), \it but not in
the source term\rm:
\be
\label{s.t}
S[J^U]=\int d^4x J^a_i \bar A^i_a,\quad \bar {\hat A}_i=
U(\hat A ^{(0)})U^{-1}.
\ee
This reflects the fact of the topological degeneration of physical fields in the  Minkowskian YM model \cite{LP1, Pervush2,David2, David1}. \par
 Generally speaking, the phase factors $U_{\cal Z}$ in  the  Minkowskian non-Abelian theory \cite{LP1, Pervush2,David2, David1},
as a "relic" of the fundamental quantization by Dirac \cite{Dir}, remember all the 
information about the chosen reference frame, vacuum BPS (Wu-Yang) monopoles, 
rising "golden section" potential (\ref{ris1}) of the instantaneous interaction between (two) total currents (\ref {total current}) and other
initial data, including their topological degeneration 
and infrared confinement (see farther) \cite{LP1}.
\par
The constraint-shell formulation distinguishes the
\it bare \rm  "gluon", as a \it weak deviation of the given vacuum monopole
with the topological index \rm $n = 0$,
and the
\it observable \rm (\it physical\rm ) "gluon"
\it averaged over the topological degeneration \rm (i.e. \it over all 
 the  Gribov copies\rm) \cite {Pervush2,Pervush3}:
\be
\label{aver}
\bar A^{phys}= \lim \sb {L\to \infty} \frac {1}{2L} \sum \sb {n=-L} ^{n=+L}\bar A^{(n)}({\bf x})
\sim \delta_{r,0};
\ee
whereas in QED any constraint-shell field is
\it a transverse photon\rm. \par 
A more detailed analysis of the latter formula will be performed  in the next section.\par
We may speak that topological Dirac variables (\ref{degeneration1}), involving the topological degeneration
of  initial states in a Minkowskian non-Abelian theory are a
physical origin of hadronization and confinement, treated as
non-local monopole effects. \par
These topological Dirac variables distinguish the
unique gauge (to within the appropriate cohomological class). \par
In QED
\it it is the Coulomb gauge\rm, 
while  in the Minkowskian YM theory \cite{LP1, Pervush2,David2, David1} \it it is 
the covariant generalization of the covariant  Coulomb gauge in
the presence of  vacuum BPS monopoles\rm. In the latter case takes place the Gribov ambiguity in the choice of the transverse Coulomb gauge \cite{Al.S., Gribov, Baal} in all the topological  sectors of the Minkowskian YM theory \cite{LP1, Pervush2,David2, David1} due to the nontrivial $U(1)\to SU(2)$ embedding. In turns, this involves the infinite number of transverse YM fields in each topological sector of that theory.  \par  
When we go over to another (relativistic) gauge of  physical sources, say $F(A)=0$, at the level of the
FP integral \cite{Fadd1}, all the monopole effects, including 
 the topological degeneration of initial data and rising potential, may be lost
(as the Coulomb potential is lost in QED in  relativistic invariant
gauges). \par  
Recall that to prove the equivalence of the Feynman integral to
the FP integral in an arbitrary gauge, we (see \cite {Pervush2}, \S2.5)
change variables:
\be
\label{change1}
A_k^* [A^F]= v[A^F](A_k^F+\partial_k)(v[A^F])^{-1}, \ee
\be
\label{change2} \psi^* [A^F]=v[A^F]\psi,
\ee
and
concentrate all the monopole effects in  phase factors
before  physical sources:
\be\label{fpis}
 S^*\;=\;\int d^4x \left( ( v[A^F] )^{-1}{\bar s}^* \psi^F +  {\bar \psi}^F  ( v[A^F] )^{-1}s^*
 +J^*_i A^*_i[A^F]\right).
 \ee 
The change of  sources removes all these
effects \cite {Pervush2}.\par
Such change of  sources was possible in the Abelian gauge theory only for
scattering amplitudes \cite {Fadd1} in  neighbourhoods
of poles of their Green functions, 
when all the particle-like excitations of fields are on theirs mass-shells
(we recommend our reader to understand this fact with the "classical" example of
electronic  propagators \cite {A.I.}). \par
However, for the cases of non-local bound states in QED and QCD and other phenomena
involving these  fields  \it off  their mass-shells\rm, the Faddeev
theorem about the
equivalence of  different "gauges"
(see,  e.g., (7.23) in \cite {Gitman}) is not valid.
\section {Free rotator: topological confinement.}
The nontrivial topology may be an origin of  the colour confinement Minkowskian non-Abelian gauge theories \it via
the complete destructive interference, in the infrared momenta region, of  the phase factors of
the topological
degeneration of initial data\rm.\par
 A good  mechanical analogy of the topological degeneration of  initial
data is the free rotator,   involving the appropriate action of a free particle
 alike to (\ref{rot}):
\be
\label{rot1}
W(N_{out},N_{in}\vert t_1)= \int \sb {0}  ^{t_1}dt \frac {\dot N ^2}{2}I;\quad p=
\dot N I,\quad H_0=\frac {p^2}{2I};
\ee
given on the ring, with the points $N(t) + n$ ($n \in \bf Z$)
being physically equivalent (cf. (\ref{PsiN})). \par
Instead of  initial data
$N(t = 0) =N_{in}$
in  mechanics,  in the space with a trivial topology,
the observer of the "topological" rotator (\ref{rot}),  (\ref{rot1})  sees \it  a   manifold of  initial data \rm
$N^{(n)}(t=0)=N_{in}+n$; $n\in \bf Z$ (we also may write down this manifold  as $N^{(n)}({\cal H}(t_0))$  in the light of the said in Section 4.3 about the Gribov  topological degeneration of non-Abelian initial data).\par
The observer does not know  where  the rotator is.
It may be in the   points $N_{in}$, $N_{in}\pm 1,N_{in}\pm 2,\dots $.
Therefore
he would  \it average \rm  the wave function (cf. (\ref{aver})):
\be
\label{Psi2}
\Psi (N)=e^{ipN}
\ee
over all the values of the topological degeneration, involving  the
$\theta$-angle measure $\exp (in\theta)$.
As a result, he gets the "observable" wave function
\be
\label{Psiob}
\Psi (N)_{observable}= \lim \sb {L\to \infty} \frac {1}{2L} \sum \sb {n=-L}^{n=L}e^{in\theta}\Psi
(N+n)=\exp\{i(2\pi k+
\theta)N\},\quad k\in \bf Z.
\ee
In the opposite case, when $p\neq 2\pi k+ \theta$,
the  appropriate wave function (i.e. \it the probability amplitude\rm)
disappears: $\Psi (N)_{observable}=0$,
\it due to the complete destructive interference\rm. \par
This corresponds to the usual
description of a repeated phenomenon in mathematical statistics (we may treat zero as the mathematical expectation value of the set 
$\bf Z$).
\par
The consequence of this topological degeneration is that the part
of  values of  momentum spectrum becomes \it unobservable \rm
in comparison with a trivial topology.\par
This fact may be treated as the \it confinement \rm  of those values  that do not coincide with
\be
\label{conf}
p_k=2\pi k+ \theta, \quad 0\leq\theta\leq \pi.
\ee
The observable spectrum also follows  from the constraint of 
equivalence of the points  $N$ and $N + 1$:
\be
\label{eqv}
\Psi (N)=e^{-i\theta}\Psi (N+1),\quad \Psi (N)=e^{ip N}. \ee
(the $\theta$-angle is, indeed, the eigenvector of the gauge transformation
$T_1 \vert \theta >=e^{i\theta}\vert \theta >$ corresponding to the
 raise of the given
topological number on  unit: $T_1 \vert n >=\vert n +1>$;
this  is valid  in the
 Euclidean as well as in the Minkowski spaces).\par
As a result, we obtain the spectral decomposition of the Green function
of the free rotator (\ref{rot1})
(as the probability amplitude of the transition from the point
$ N_{in}$ to the point $N_{out}$) over the observable values of the spectrum (\ref{conf}):
\be
\label{ampl}
G(N_{out},N_{in}|t_1)\equiv <N_{out}|\exp(-i\hat Ht_1)|N_{in}>
 = \frac{1}{2\pi}\sum\limits_{k=-\infty }^{k=+\infty }\exp\left[
 -i\frac{p_k^2}{2I}t_1+ip_k(N_{out}-N_{in})\right]~.
\ee
Using the connection with the Jacobian theta-functions \cite {Pollard}:
\be
\label{Theta}
\Theta_3(Z\vert \tau)= \sum \sb {k=-\infty} ^{k=\infty}\exp [i\pi k^2\tau+2ikZ]=  (-i\tau)^{-1/2}\exp [\frac {Z^2}{i\pi\tau}]
\Theta_3(\frac {Z}{\tau}\vert -\frac {1}{\tau}),
\ee
we may represent the expression (\ref{ampl}) as the  sum over all the paths:
\be \label{paths}
G (N_{out},N_{in}\vert t_1)=\sqrt{\frac {I}{4\pi i t_1}} \sum \sb {n=-\infty} ^{n=\infty}
\exp [i\theta n]\exp [+iW(N_{out},N_{in}\vert t_1)
)],
\ee
with
\be
\label{w}
W (N_{out}+n,N_{in}\vert t_1)=\frac {(N_{out}+n-N_{in})^2 I}{2t_1}
\ee
being the rotator action (\ref{rot1}).
\section {Infrared quark confinement as a destructive interference of "large" Gribov topological multipliers.}
The topological confinement similar to the complete destructive interference
of the phase factors of the topological degeneration (\it the latter one is  a pure quantum effect\rm) may  occur in classical non-Abelian
field theories. \par
Recall that, at the time of the  first paper by Dirac \cite {Dir},
the so-called "classical relativistic field theories" were found in
the papers by Schr\"odinger, Fock, Klein, Weyl \cite{Fock,Weyl} as  patterns of
\it relativistic quantum mechanics\rm,
i.e. as  results of the primary quantization.
The phases of the gauge transformations were introduced by
Weyl \cite {Weyl} \it as pure quantum magnitudes\rm.\par
The  free rotator theory   implies that the topological degeneration may be
removed  as all the Green functions are averaged over the values of the
topological
variable and all  the possible angles of orientation of the  monopole unit
vector $\bf n$ 
in the group space (instead of the instanton averaging over
interpolations \cite{Bel} between  different vacua
in the Euclidean space).\par
Averaging over all the parameters of  degeneration may lead to the
complete destructive interference of all the colour amplitudes
\cite {Pervush3,Ilieva,Nguyen,Azimov}. \par
To verify the latter statement, we should recall that  the Gribov  matrices $v^{(n)} ({\bf x})$ in  (\ref{degeneration1}), (\ref{degeneration}) have, indeed, the spatial asymptotic $v^{(n)} ({\bf x})\to 1$ as  $\vert {\bf x}\vert \to \infty$, (\ref{bondari}) \cite {Pervush1,Azimov}.\par
In effect, (\ref{bondari})  found to be the normalization condition  that would be imposed onto the Gribov  matrices $v^{(n)} ({\bf x})$ in the Minkowskian YM theory \cite{LP1, Pervush2,David2, David1} to ensure a good  infrared behaviour of this theory. \par
In Section 4.3 we  have ascertained the necessary look \cite {Azimov} of Gribov matrices $v^{(n)} ({\bf x})$ for satisfying the normalization (\ref{bondari}). As a result, we came to the spatial asymptotic (\ref{newfact})- (\ref{eiler})  for Gribov matrices $v^{(n)} ({\bf x})$ as $\vert {\bf x}\vert \to \infty$.  \par 
Now our task  is to demonstrate that "large" Gribov matrices $v^{(n)} ({\bf x})$, possessing the  spatial asymptotic (\ref{newfact})- (\ref{eiler}),  disappear, in the infrared limit $\vert {\bf x}\vert \to \infty$, in quark Green functions in all the orders  of the perturbations theory. 
\par 
The said quark Green functions may be obtained issuing from the FP path integrals formalism \cite {Azimov}. \par 
Let us  fix a gauge, say  
\be \label {FP2}
{\rm det}~ \hat \Delta =0,
\ee
for the FP determinant ${\rm det}~ \hat \Delta $ in (\ref{sigma}). \par 
It is highly obvious (see, e.g., the work \cite {Baal}) that in the Minkowskian YM model \cite{LP1, Pervush2,David2, David1} the gauge  (\ref{FP2}) is mathematically equivalent to the Gribiov ambiguity equation (\ref{Gribov.eq}).
\par
The standard FP path integral \cite {Fadd1}, as it is well known,  has the look \cite {Pervush2}
\begin{eqnarray}\label{fpi}
 Z^{*}[s^*, {\bar s}^*, J^*]\;=\;\int \prod_{\mu}DA^F_\mu D
 \psi^F D{\bar \psi}^F \Delta _{FP}^F
 \delta (F(A^F)) e^{iW[ A^F,\psi^F, {\bar \psi}^F ] + S^*},
 \end{eqnarray}
involving the source term (\ref{fpis}). \par
In particular, the gauge (\ref{FP2}), specifying, in the Minkowskian YM theory \cite{LP1, Pervush2,David2, David1}, the Gribiov ambiguity in the choose of transverse YM fields: topological Dirac variables  (\ref{degeneration1}), implies the following look for the FP path integral (\ref{fpi}) \cite{Gitman,Azimov}:
$$
Z_{R,T} (s^*, \bar s^*,J^*)= \int D A_i^* D\psi^* d \bar\psi^* ~{\rm det}~ \hat \Delta ~\delta (\int \limits_ {t_0}^t d\bar t D_i (A)\partial_0 A^i)
$$
\be \label{Zr}
\times
\exp \{ i \int \limits_ {-T/2}^ {T/2} dt \int \limits_ {\vert {\bf x}\vert \leq R} d^3x [{\cal L} ^I (A^*, \psi^*) +  \bar s^*\psi^* +\bar \psi^* s^* +J^{*a}_i A^{i*}_a]\}. 
\ee 
In this formula the Lagrangian density ${\cal L}^I$ corresponds to the constraint-shall action of the Minkowskian YM theory (or Minkowskian QCD) \cite{LP1, Pervush2,David2, David1} taking on the surface of the Gauss law constraint (\ref{Gauss}); $R$ is an enough large real number, we may think  that $R\to \infty$. \par
This involves the formal explicit dependence  of the FP path integral (\ref{Zr}) on Dirac variables $A^*$, $\psi^*$, $\bar\psi^*$. \par
The same cause implies the gauge fixation $\delta (\int \limits_ {t_0}^t d\bar t D_i (A)\partial_0 A^i)$ in (\ref{Zr}). This gauge is mathematically equivalent to the Coulomb transverse gauge (\ref{Aparallel}), (\ref{transv}). \par
At the spatial infinity, $\vert {\bf x}\vert \to \infty$ (i.e. in the infrared region of the momenta space), the general formulas (\ref{change1}), (\ref{change2})  \cite{Pervush2} for (topological) Dirac variables inherent in the Minkowskian YM theory \cite{LP1, Pervush2,David2, David1} acquire the asymptotical  look (\ref{newfact}) \cite{Azimov}.\par
Upon this substituting and  averaging  the FP path integral (\ref{Zr}) over the Gribov topological degeneration of initial data, i.e. over the set $\bf Z$ of integers, we get the generating functional that corresponds to the topological confinement in the Minkowskian YM theory \cite{LP1, Pervush2,David2, David1}:
\be \label{Zcon}
Z_{conf} (s^*, \bar s^*,J^*)= \lim _{\vert {\bf x}\vert\to \infty,~T\to \infty} \lim _{L\to \infty} \frac{1}{L} \sum \limits _{n=-L/2}^{n=L/2}  Z_{R,T}^I (s^*_{n,\phi_i}, \bar s^*_{n,\phi_i},J^*_{n,\phi_i}), \ee
with $ Z_{R,T}^I (s^*_{n,\phi_i}, \bar s^*_{n,\phi_i},J^*_{n,\phi_i})$ being the FP path integral (\ref{Zr}) rewritten in terms of Gribov exponential multipliers (\ref{newfact}) \cite{Azimov} at the spatial infinity.\par
It is worth to note a good consent between the generating functional (\ref{Zcon}) and  Eq. (\ref {aver})  \cite{Pervush2} for an observable (physical) gluon.
\par
The variation of this generating functional by the sources:
$$ (\prod \limits _{\alpha=1}^3 \frac{\delta}{\delta s^*_{n,\phi_\alpha}})(\prod \limits _{\beta=1}^3 \frac{\delta}{\delta \bar s^*_{n,\phi_\beta }})(\prod \limits _{\gamma=1}^3 \frac{\delta}{\delta J^*_{n,\phi_\gamma }}),$$
is accompanied by  averaging  the appropriate Green functions over the Euler angles $\phi_i$ ($i=1,2,3$) in (\ref{ln}). The latter  ones  describe the positions of colour degrees of freedom  in the $\{\phi_i \}$ reference frame. Herewith the Euler angles $\phi_i$ would satisfy the normalization 
(\ref{fas.ysl-e}) to achieve the quested asymptotic (\ref{bondari}) for Gribov exponential multipliers $ v^{(n)} ({\bf x})$.\par
Let as compute the one-particle Green function for a quark in   Minkowskian YM QCD \cite{LP1, Pervush2,David2, David1} involving fermionic degrees of freedom including as perturbation fields over the Minkowskian YM vacuum \cite{David1}.\par
In the lowest order of the perturbation theory we then get 
\cite{Nguyen, Azimov} 
\be \label{oneqw}
G({\bf x},{\bf y})= \frac {\delta}{\delta s^*(x)} \frac {\delta}{\delta \bar
s^*(y)} Z_{conf} (s^*, \bar s^*,J^*)\vert_{ s^*= \bar s^*=0} =G_0(x-y) f({\bf x},{\bf y}),
\ee
with $ G_0(x-y)$ being  the quark propagator in the perturbations theory and
\be \label{interf}
f({\bf x},{\bf y})= \lim_{\vert {\bf x}\vert \to \infty, ~ \vert {\bf y}\vert \to \infty} \lim_{L \to \infty} (1/L) \sum \limits _{n=-L/2}^{n=L/2} v^{(n)} ({\bf x}) v^{(n)} ({\bf -y}). \ee 
To estimate the function $ f({\bf x},{\bf y})$ \cite{Azimov}, we should substitute the asymptotic expression  (\ref{newfact}) for Gribov exponential multipliers in the latter formula. \par  
This results $ f({\bf x},{\bf y})=1$ due to the spatial asymptotic $v^{(n)} ({\bf x})\to 1 $ {\rm as}  $\vert {\bf x}\vert \to \infty$ for the Gribov topological multipliers (\ref{newfact}), involving the normalization 
(\ref {fas.ysl-e}).\par  
Thus we see that only "small" Gribov exponential multipliers  $v^{(0)} ({\bf x})$ contribute in $ f({\bf x},{\bf y})$ and, therefore, in the one-particle quark Green function (\ref{oneqw}).\par
The same result  may be also demonstrated \cite{Azimov} in the momenta space.  Upon  substituting
$$ t^0=0;\quad t^a= \frac{\tau ^a}{r}\pi;\quad a=1,2,3;  $$
in   (\ref{newfact}) we get  
\be \label{Gp}
G(p)= \lim_{L \to \infty} (1/L) \sum \limits _{n=-L/2}^{n=L/2} \frac{1}{\hat p+ \hat t n}=0 \quad   (\hat p= p_\mu \gamma ^\mu)
\ee  
for the one-particle quark Green function in the momenta space.  It becomes $O(1/p)$ in the limit $n\to \infty$.   This confirms Eq. (\ref{interf}) valid in the coordinate space. \par
Consider now \cite{Azimov} a quark loop, treated as the vacuum expectation value of fermionic currents
\be \label{qloop} 
<j^\Gamma(x), j^\Gamma(y)>,  \quad j^\Gamma= \bar \psi \Gamma \psi. \ee 
In the lowest order of the perturbation theory we then get 
$$
 <j^\Gamma(x), j^\Gamma(y)>_p =\int d\Omega (\phi_x) d\Omega (\phi_y) \lim_{L \to \infty} (1/L) \sum \limits_ {n=-L/2}^{n=L/2} {\rm tr}~ [v_ {n, \phi_x }^{-1} ({\bf x}) \times
$$
\be \label{interf1}
\times
\Gamma~  v_ {n, \phi_x } ({\bf x}) G_0(x-y) v_ {n, \phi_y }^{-1} ({\bf y}) \Gamma ~ v_ {n, \phi_y } ({\bf y}) G_0(y-x)],
\ee 
with $ d\Omega $ being the integral by Euler angles and $\Gamma$ being a combination of Dirac matrices. We apply in the latter formula the denotation $ v_ {n, \phi_x } ({\bf x})$ for  Gribov exponential multipliers (\ref{newfact}). This points out to their transparent connection with the Euler angles $\phi_i$ ($i=1,2,3$). \par
When a matrix  $\Gamma$ is a colour scalar: $v^{-1}  \Gamma v= \Gamma$, we deal with a quark loop with ordinary free propagators:
\be \label{frprop}
<j^\Gamma(x), j^\Gamma(y)>_p = {\rm tr} ~[\Gamma G_0(x-y) \Gamma G_0(y-x)],
\ee 
whose imaginary part is different from zero. \par
But when  a matrix  $\Gamma$ is "coloured", we come to expressions of the (\ref{oneqw})- (\ref {Gp}) type for quark loops. \par 
For example, instead of the  expression (\ref {Gp})  for the one-particle quark Green function in the momentum representation, we then get for a  quark loop the expression
\be \label{petlja} \Pi (q)= \lim_{L \to \infty} (1/L) \sum \limits_ {n=-L/2}^{n=L/2} \int d^4x  {\rm tr} [\Gamma G_0(p+tn) \Gamma G_0(q-(p+tn))]= \int d^4x  {\rm tr} [\Gamma G_0(p) \Gamma G_0(q-p)],
\ee 
in which takes place the complete cancellation of infinite-large  momenta. \par 
The alike topological confinement also takes place, in the Minkowskian  QCD \cite{LP1, Pervush2,David2, David1} for YM (gluonic) fields.\par
As a result, only topological trivial gluons (\ref{aver}) \cite{Pervush2} and topological trivial quarks (involving "small" Gribov multipliers $ v^{(0)} ({\bf x})$ in (\ref{change2})) survive in the  infrared momenta region. Just  they form (as we shall demonstrate below) the observable (physical)  hadronic bound states. 
\par
The similar mechanism of the infrared confinement was described for the \linebreak 
two-dimensional model in the works \cite{Hooft4,Ebert1}. \par
Upon  averaging by infrared parameters of the Gribov topological degeneration, disappear all the quark Green functions which are not scalars under colour gauge transformations of the (\ref{degeneration1}) type. type. On the other hand, colourless Green functions coincide with Green functions in ordinary perturbation QCD.  They are similar to correlators between electromagnetic and weak currents.  Thus we come to the quark confinement a la naive partonic model by Feynman \cite{Feynman}.\par
In this case only \it colourless \rm ("hadronic")
states \it form the complete set of physical states\rm.
Using the example
of a free rotator, (\ref{rot1}),  we see that the disappearance of the part of
physical states due to the infrared topological confinement \cite{Azimov} does not
violate the composition law for a Green function:
\be
\label{comp}
G_{ij}(t_1,t_3)= \sum \sb {h} G_{ih}(t_1,t_2)G_{hj}(t_2,t_3),
\ee
that is  specified as  the probability amplitude to  find the system described by the 
Hamiltonian $H$ in the state $j$ in the  time instant $t_3$  when in the  time instant $t_1$
this system was in the state $i$, where $(i; j)$ belong to the complete set
of all the physical states $\{h\}$:
\be
\label{compl.set}
G_{ij}(t_1,t_3)=<i\vert \exp -i \int \sb {t_1}^{t_3} H)\vert j>.
\ee
A particular case of this composition law (\ref{comp}) is the unitarity of the  QCD S-matrix:
\be
\label{un}
SS^+ = I \Longrightarrow  \sum \sb {h} <i\vert S \vert h><h \vert S^+ \vert j>= <i\vert j>,
\ee
known as the law of the probability conservation for the  S-matrix elements ($S = I + iT$), where
\be
\label{cons}
\sum \sb {h} <i\vert T \vert h><h \vert T^* \vert j>= 2{\rm Im}~<i\vert T \vert j>
\ee
(cf.  (64.2), (71.2) in \cite {BLP}).\par
The left-hand side of this law is similar to the spectral series of the
free rotator, (\ref{ampl}).
Herewith the infrared destructive interference \cite {Azimov} of Gribov "large" multipliers $v^{(n)}({\bf x})$ ($n \neq 0$) keeps only the colourless "hadronic" states. \par
 The right-hand side of the probability conservation law (\ref{cons}), far from resonances, may
be represented by the perturbation series over the Feynman diagrams
that follow from the Hamiltonian. \par
Due to the manifest gauge invariance of the fermionic Hamiltonian (Lagrangian) in (\ref{fermioni}) \cite{David1}, 
\be
\label{invham} 
H[A^{(n)},q^{(n)}]=H[A^{(0)},q^{(0)}],
\ee
with $q$ are being  fermionic (quark) degrees
of freedom. Herewith the record $q^{(n)}$  for quark degrees of freedom implies that there also are  Dirac variables having the general  look (\ref {change2}). \par
The said implies  that the Hamiltonian $H[A^{(0)},q^{(0)}]$ is  invariant
with respect to large gauge transformations (\ref{degeneration}) turning YM fields into physical topological Dirac variables. \par
Thus the Minkowskian QCD  Hamiltonian does not depend on  Gribov large multipliers $v^{(n)}({\bf x})$ ($n\neq 0$) given in  (\ref{mon.deg}). We may treat this  as  \it the  complete destructive interference of Gribov  factors of the topological degeneration in the QCD Hamiltonian\rm.  \par
On the  other hand, we come to
the usual treatment of the colours confinement in QCD, when the
colourless fermionic degrees of freedom $q^{(0)}$ are considered as
"\it hadronic\rm"  states,  which we may identify with the Feynman partons
\cite{Feynman}.\par
The considered above holonomies 
theory (\ref{bg})- (\ref{coh}) allow us to draw the conclusion that the colour
confinement in the Minkowskian YM theory \cite{LP1, Pervush2,David2, David1},  involving the Gribov equation  (\ref{Gribov.eq})
 and its vacuum solution  (\ref{phase}), is determined by the restricted
holonomies group  $\Phi^0$  generated by the  zero topological 
sector of this YM theory \rm (more precisely, by the YM fields $A^{(0)}$ of this
sector satisfied the Coulomb gauge (\ref {Aparallel}) and boundary conditions (\ref {bondari}) at the spatial infinity) and representing the "small" $U(1)$ gauge transformations).
We may interpret this as the  confinement
criterion in Minkowskian YM QCD \cite{LP1, Pervush2,David2, David1} \rm (that is also  correct for the gluonic theory involving the
$SU(3)_{col}\to SU(2)\to U(1)$ spontaneous breakdown). \par
It turns out that (closely entangled each with other) the topological and quark confinements inherent in the Minkowskian model  \cite{LP1, Pervush2,David2, David1} may be correct described in terms of the mixed task (see, e.g., \S 4.1 to Chapter 1 in \cite{Vlad}) to the Gribov ambiguity equation (\ref{Gribov.eq}): the second-order differential equation in partial derivatives. \par
The said implies that we should supplement the initial Cauchy conditions (\ref{Aparallel}), (\ref{init}): in the zero topological sector, to the Gribov  equation (\ref{Gribov.eq}) (in the initial time instant $t_0$) by the boundary  condition (\ref{bondari}) at the spatial infinity, involving the asymptotical look (\ref{newfact}) \cite{Azimov} for Gribov topological multipliers $v^{(n)}({\bf x})$. \par
It is highly appropriately to note here that the said mixed task for the Gribov  equation (\ref{Gribov.eq}) may found to be somewhat unexpected for our readers who  devotes themselves to  differential equation in partial derivatives.  \par
 It is well known that elliptical differential equation in partial derivatives, to which belongs the  Gribov  equation (\ref{Gribov.eq}), need in  none initial  conditions:  there are purely stationary differential equation  that do not depend on the  time $t$ (see, e.g., \S 4.1 to Chapter 1 in \cite{Vlad}). \par 
But  in the Minkowskian non-Abelian theory  \cite{LP1, Pervush2,David2, David1} there is, for all that, the indirect dependence  of the  Gribov  equation (\ref{Gribov.eq}) on the  time $t$. \par
This is associated with  considering the Minkowskian  non-Abelian theory \cite{LP1, Pervush2,David2, David1}  on the surface  of the  YM Gauss law constraint (\ref{Gauss}). In  this case the Dirac removal \cite{Dir} of temporal YM components involves the  Gauss law constraint (\ref{Gauss2}), which we then resolve in the Coulomb transverse gauge 
(\ref{cshg}), (\ref{Aparallel}).\par
On the other hand, just the ambiguity in the choose of the  Coulomb transverse gauge (\ref{Aparallel}) is described by the  Gribov  equation (\ref{Gribov.eq})  \cite{Al.S.,Baal}. The origin of this ambiguity in the definition of Dirac (topological) variables (\ref{degeneration}) apart from  Gribov stationary multipliers $v^{(n)}({\bf x})$ \cite{LP1}.\par
Thus the both things:  the  YM Gauss law constraint (\ref{Gauss}), (\ref{cshg}) and the  Gribov  ambiguity equation (\ref{Gribov.eq}), are closely related   each with other. \par 
In effect, the appearance of the initial Cauchy conditions (\ref{Aparallel}), (\ref{init}) to the elliptical Gribov  equation (\ref{Gribov.eq}) is an original trace of the  Minkowski space-time in the non-Abelian model \cite{LP1, Pervush2,David2, David1}. This involves the consideration of that  model on the  space-like surface ${\cal H} (t_0)$ in the   Minkowski space-time (in the initial time instant $ t_0$) where the Gribov topological degeneration of initial YM data occurs. \par 
The QCD Hamiltonian $H$  contains the perturbation series in terms only of
zero degree of the map fields  (i.e. in terms of  \it constituent colour
particles\rm) that may be identified \it with  Feynman partons \rm \cite{Feynman}.\par
The Feynman path integral of the  (\ref{N.i}) type, as the generating
functional of these perturbations series, is an analogue of the sum
over all the paths of the free
rotator (\ref{paths}).\par
Therefore confinement, in the spirit of the complete destructive
interference of  colour amplitudes \cite {Pervush2,Pervush3,Nguyen,Azimov},
and the law of the probability conservation for the S-matrix, (\ref{cons}),
imply the \it Feynman quark-hadronic duality \rm  that is the base of all the
partonic models \cite{Feynman} and  QCD applications \cite{Efremov}.\par
 The quark-hadronic duality gives the method of  a direct experimental measurement
 of  quark and gluonic quantum numbers issuing from  the given deep-inelastic 
scattering cross-section \cite {Feynman}.\par
For example, according to Particle Data Group,
the ratio of  the sum of the probabilities of the $\tau$-decay
 hadronic  modes to the probability of the $\tau$-decay muonic  mode is
\be
\label{ratio}
\frac {\sum \sb {h} w_{\tau \to h}}{w_{\tau \to \mu}}=3.3 \pm 0.3 .
\ee
This is the left-hand side of  Eq. (\ref{cons})
normalized to the value of the leptonic  mode probability of the $\tau$-decay.
On the right-hand side of  Eq. (\ref{cons}) we have the ratio of the imaginary part
of  the sum over  the quark-gluonic diagrams (in terms of constituent
fields free from  Gribov phase factors $v^{(n)}({\bf x})$) to the one of the leptonic
diagrams. In the lowest order of  perturbative QCD, on the right-hand
side we get the number of colours $N_c$, therefore
\be
\label{Nc}
3.3 \pm 0,3 = N_c.
\ee
Thus the degeneration of initial data in Minkowskian  YM QCD \cite{LP1, Pervush2,David2, David1}   may explain  not only
"\it why we do not see  quarks\rm",
but also "\it why we may measure their
quantum numbers\rm". \par
The considered mechanism of confinement in Minkowskian  YM QCD \cite{LP1, Pervush2,David2, David1}, due to a quantum interference of the
phase factors of the topological degeneration,  disappears  upon a change of the
"physical" sources:  $A^*J^*\Longrightarrow A J$,
called the transition to  another gauge in the gauge-fixing method \cite{Fadd1}.
\par
This  mechanism is a highly delicate thing, based (mathematically) on the just described mixed task (\ref{Gribov.eq}) involving the initial conditions (\ref{Aparallel}), (\ref{init}) and asymptotical boundary condition (\ref{bondari}) at the spatial infinity. \par
 Thus it becomes obvious that the transition to  another gauge from the transverse Coulomb one, (\ref{Aparallel}), destroys  the said mechanism of confinement in Minkowskian  YM QCD \cite{LP1, Pervush2,David2, David1}.\par
Instead of the hadronization and confinement, we obtain then, in the "relativistic" FP integral (\ref{fpi}),
only  scattering amplitudes  of  free partons. But these 
amplitudes do not exist as physical observables in the Dirac
quantization scheme depending on initial data. \par
Removing (topological)  Dirac variables (\ref{degeneration1}) (that are transverse and gauge invariant) via going over to  another gauge, we herewith  also remove the dependence of the  FP integral (\ref{fpi}) on a reference frame and initial data \cite{Pervush2}. \par
To understand the latter statement we should recall that Dirac variables  are manifest relativistic covariant\rm. \par 
A good analysis of this feature of Dirac variables was carried out in the Polubarinov review \cite{Polubarinov} (see also Section 2.3 in \cite{Pervush2}).  \par
Thus going over to an arbitrary gauge (that is not transverse one) in the gauge-fixing method \cite{Fadd1}, we also remove the dependence of the  gauge theory obtained in  such a wise on a  reference frame and initial data.  \par 
The said confirms the warning made by Schwinger \cite{Schwinger} 
that gauges that
are independent of a reference frame may be physically
inadequate to the fundamental operator quantization;
i.e.  they may distort the spectrum of the original
system
\footnote{"We reject all Lorentz gauge  formulations as unsuited to the role of providing  the fundamental operator quantization" \cite{Schwinger}. 
}.
\section { U(1)-problem.}
 The  value of the vacuum chromo-magnetic field $<B^2>$ may be estimated
 by the description of a process involving ABJ  anomalies.\par
 The simplest process of such  type is the interaction
 of a pseudo-scalar bound state with an ABJ anomaly.\par
The potential source of ABJ  anomalies in any QCD model are axial currents that follows, due to the variation principle, from the QCD Lagrangian, that always may be recast in such a wise to contain pseudoscalar combinations of fermionic fields involving the $\gamma_5$ Dirac matrices. \par  
We may recommend our readers several examples haw  to transform  the QCD Lagrangian to result pseudoscalar combinations of fermionic fields. Such patterns of the appropriate transformations were demonstrated in the monograph \cite{Cheng}, in \S 10.2, and in the papers \cite{Ebert,Volkov,Egu}.\par  
In particular, axial currents may enter, as an item, the total current  (\ref{total current}) \cite{David1}, and this implies their contribution in the instantaneous current-current  interaction (\ref{cint}). \par
Neglecting  all the mesonic channels excepting $\eta_0$ one, with \cite{David3}
\be \label {eta0}
\eta_0= (u\bar u+ d\bar d +s \bar s)/ \sqrt {3},
\ee 
one may incorporate the anomalous item \cite{David2,David1}
\be\label{ven}
\tilde W_{\rm anomaly}^{\eta_0}[\eta,\bar \eta]=C_{\eta}\int dt
\bar \eta(t,0) I_c\gamma_5\eta(t,0) \frac{g^2}{16 \pi^2}
\int d^3x F^a_{\mu \nu}{}^*F^a_{\mu \nu},
\ee
in the general (Minkowskian) QCD action. \par
Here $\eta$, $\bar \eta$ are the fermionic  sources, $C_\eta$ is a constant  we shall specify below. \par
At the level of Feynman diagrams this anomalous action corresponds \cite{David3} to the interaction between a pseudoscalar glueball field
\be\label{glshar}
{\cal Q}=\gamma_5 F^a_{\mu \nu}{}^*F^a_{\mu \nu}
\ee
and  two  $\eta_0$-meson scalar states   (\ref {eta0}) attached to  this glueball. \par
Due to (\ref {winding num.}),
\be \label{nor}
\frac{g^2}{16 \pi^2}
\int d^3x F^a_{\mu \nu}{}^*F^a_{\mu \nu}= \dot N(t).
\ee 
Besides that, following the work \cite{Arsen}, we utilize the normalization
\be \label{nor1}
\frac{g^2}{8 \pi^2}\int d^3D_i^{ab}({\Phi})\Phi^b_0 B_i^a({\Phi})=1
\ee
for the vacuum magnetic field $\bf B$, specified by the Bogomol'nyi equation (\ref {Bog}). \par
The physical anomalous  term $\tilde W_{\rm anomaly}$, (\ref{ven}), in the
$\eta_0$-meson channel  takes the look \cite{Veneziano}
\be\label{ven1}
\tilde W_{\rm anomaly}^{\eta_0}[\eta,\bar \eta]=
C_{\eta}\int dt \eta_0 \dot N,~~~
\eta_0(t)=\bar \eta(t) I_c\gamma_5\eta(t).
\ee
Following Veneziano \cite{Veneziano}, we may identify $\eta_0(t)$ with the field of the
$\eta_0$-meson (\ref{eta0}) at rest (multiplied onto $\gamma_5$). The effective Minkowskian QCD action including the anomaly term
(\ref{ven}) and the free rotator  term, (\ref{rot}), is then \cite{David1}
\be \label{eta}
W_{\rm eff}=\int dt [\frac{\dot N^2 { I}}{2}+ \eta_0C_{\eta}\dot N +\frac{1}{2} \dot \eta_0 ^2V - 
\frac{\eta_0^2 m_0^2 V}{2}],
\ee
with $m_0$ being the standard current quark mass contribution to the 
$\eta_0$-meson mass. \par
The effective action (\ref{eta}) is a particular case of the universal effective  action inherent in
gauge theories for  descriptions of  anomalous interactions \cite{LP1,Pervush2,David2}:
\be
\label{inter}
 W_{eff}=\int dt \left\{\frac 1 2 \left({\dot\eta_M}^2-M_P^2{\eta_M}^2\right)
 V +
 C_M\eta_P \dot X[A^{(N)}] \right\},
\ee
 with $\eta_M$ being a bound state with the mass $M_P$ in its rest reference frame and
 $X[A^{(N)}]$ being the topological  "winding number" functional. \par
 In three-dimensional QED${}_{(3+1)}$ this action,   involving the constant \cite {Pervush2}
 \be
\label{C}
 C_M=C_{\rm positronium}=\frac{\sqrt{2}}{m_e}8{\pi}^2
 \left(\frac{\underline{\psi}_{Sch}(0)}{m_e^{3/2}}\right),
 \ee
 describes the decay of a positronium $\eta_M=\eta_P$ into two photons associated with the 
 "winding number" functional
 \be
\label{wn}
 \dot X_{ QED}[A]= \frac{e^2}{16 \pi^2}
 \int d^3x F_{\mu \nu}{}^*F^{\mu \nu}\equiv\frac{e^2}{8 \pi^2}
 \int d^3x \varepsilon_{ijk}\dot A^i(\partial^jA^k- \partial^kA^j)~.
 \ee
 In one-dimensional QED${}_{(1+1)}$ the action (\ref{inter}),  involving the constant
 $C_M=2\sqrt{\pi}$ and the "winding number" functional
 \be
\label{funct}
 \dot X_{ QED}(A^{(N)})=\frac{e}{4\pi}
 \int\limits_{-V/2 }^{V/2 }dx F_{\mu\nu}
 \epsilon^{\mu\nu}= \dot N(t)~\Rightarrow~~F_{01}=\frac{2\pi\dot N}{eV},
  \ee  describes the  mass of the Schwinger
 bound state $\eta_P=\eta_{Sch}$  when the action (\ref{inter})
 is added by the action of the Coleman electric field \cite{Ilieva,Gogilidze}:
 \be
\label{el.f}
  W_{ QED} =\frac{1}{2}\int dt  \int\limits_{-V/2 }^{V/2 }dx F^2_{01}=
 \int dt\frac{\dot N^2 I_{ QED}}{2},
 \ee
 where
 \be
\label{IQED}
 I_{ QED}=\left(\frac{2\pi}{e}\right)^2\frac{1}{V}.
  \ee
 It is easy to see that the diagonalization of the total Lagrangian
 of the 
 \be
\label{Lag}
 L=[\frac{\dot N^2I}{2}+C_M\eta_M \dot N] =
 [\frac{(\dot N+C_M\eta_M/I)^2I}{2}- \frac{C_M^2}{2IV} \eta_M^2V ]
 \ee
 type implies  the mass of a pseudo-scalar meson in
  QED${}_{(1+1)}$:
\be
\label{mQED} 
  \triangle {M}^2=\frac{C_{M}^2}{I V}=\frac{e^2}{\pi}~.
\ee
In  Minkowskian QCD${}_{(3+1)}$ \cite{LP1, Pervush2,David2,David1} the diagonalization procedure similar to 
  (\ref{Lag}) implies  an additional mass of the
   $\eta_0$ meson:  $$
  L_{eff}= \frac{1}{2}[{\dot\eta_0}^2-\eta_0^2(t)(m_0^2
 +\triangle m_{\eta}^2)]V~,
 $$
 $$
  \triangle {m_\eta}^2=\frac{C_{\eta}^2}{I_{ QCD}V}
 = \frac{N_f^2}{F_{\pi}^2}\frac{\alpha_s^2<B^2>}{2\pi^3},
 $$ 
with 
\be
\label{mQCD}
 I_{ QCD}\equiv I=\left(\frac{2\pi}{\alpha_s}\right)^2\frac{1}{V<B^2>}  \ee  
founded from (\ref{I}) \cite{David2}. 
\par  
 This result allows us to estimate the value of
  the vacuum chromomagnetic field in Minkowskian QCD${}_{(3+1)}$ \cite{LP1}:
\be
\label{Bav} 
 <B^2>=\frac{}{}\frac{2\pi^3F_{\pi}^2\triangle
 {m_\eta}^2}{N_f^2\alpha_s^2}=\frac{0.06 GeV^4}{\alpha_s^2}
 \ee 
 by
using  estimating $\alpha_{s}(q^2 \sim 0)\sim 0.24$  
\cite{David1,Bogolubskaja}. \par
Upon the computations we may remove
  the infrared regularization $V \to \infty $.
\section* {Conclusion.}
 The principal problems of the discussion about stable
 vacuum states in any non-Abelian theory are the classes of functions
 and singularities. \par
 These problems exist in all models of the QCD (YM) vacuum, including
 instantons \cite{Bel,Al.S., Ryder, Cheng, Gribov} described by the 
delta-function-like singularities in the
 Euclidean space $E_4$.\par
 Mysteries of  nature are not only the actions and symmetries, but
 also the class of functions  involving finite energies densities, applied  in QFT (including QED)
 for  description of physical processes.\par
 If we explain any effect by these  singularities,
 choosing a model of the nontrivial QCD (YM) vacuum, we should answer the
 questions: "Where are singularities of this vacuum from?" and
 "What is a physical origin of these singularities?".
 \par
 We have  presented  here the model \cite{LP1, Pervush2,David2,David1} of the physical vacuum in the  YM (gluonic)
 theory in the monopole class of functions  involving the finite energies densities and
 without any singularity in a finite volume 
\footnote{ Excluding, in effect, the infinite narrow cylinder with the diameter $O(\epsilon)$ around the axis $z$ in the chosen rest reference frame where Higgs vacuum BPS monopoles disappear \cite{Pervush2, David1}, while the vacuum "magnetic" field $\bf B$ becomes infinite \cite{BPS}. },
 as  a consequence ("smile") of the scalar Higgs field that
 disappears (like the Cheshire cat) from the
 spectrum of physical excitations of the considered theory in the limit of the
 infinite spatial volume. \par
In other words,
 we have  demonstrated that there exists a mathematically correct model of the YM (gluonic)
vacuum,  involving the
 finite physical energy-momentum spectrum in the Minkowski space, underlain  by the  Bose condensate of the Higgs scalar field  
 in the limit of its infinite mass $m/\sqrt{\lambda}$ (as the spatial volume $V\to \infty$).\par
 The initial $SU(2)$ gauge symmetry  in  the Minkowskian
 YM  theory \cite{LP1, Pervush2,David2,David1} 
  is spontaneously broken down. This 
 $SU(2)\to U(1)$ breakdown always occurs 
in the presence of  the Higgs $SU(2)$ isovector. \par 
 As the Higgs field goes to the statistical (vacuum) expectation value
 at the spatial infinity, this implies  the nontrivial
 topological structure of the residual   symmetry group, $U(1)$,
 induced by the Higgs vacuum expectation value. \par 
  This nontrivial topological structure implies
 the presence of  topological (magnetic) charges in this theory, i.e.  \it the inevitability of the monopole configurations of the Minkowskian
YM vacuum with
 finite energies densities\rm.\par
 We have considered the here presented theory in the BPS limit (\ref{lim}) \cite{Al.S.,BPS} when the self-interaction
between the Higgs particles goes to zero. This allows us to consider 
Higgs particles (in the limit of  their infinite number, i.e. at the
level of  statistical physics) as \it an ideal gas\rm.
We impose an additional condition to be stationary on
this ideal gas (Bose condensate).
This choice influences, in the end, the stationary nature of the monopole
configurations of the YM vacuum. \par
Although  the self-interaction
between the Higgs particles goes to zero in the BPS limit (\ref{lim}), the system of YM and Higgs fields, actually present in the Minkowskian
YM model \cite{LP1, Pervush2,David2,David1}, is, indeed, a system of fields with the strong coupling determining by the YM (gluonic)  coupling constant $g$.\par
This stipulates, in effect,  all the features inherent in the Minkowskian
YM model \cite{LP1, Pervush2,David2,David1}, via the  Bogomol'nyi equation (\ref{Bog}), (\ref{Bog1}) depending on the coupling constant $g$. \par
As we have noted at the beginning of our discussion, the vacuum of such model, involving the  strong coupling of gauge fields, is similar to the superfluid component in a helium II \cite{N.N.}. As in that case, long-range correlations of local excitations and cooperative degrees of freedom \cite{Pervush1} also appear in the Minkowskian
YM theory \cite{LP1, Pervush2,David2,David1}.
\par
The Bogomol'nyi equation (\ref{Bog}), obtained at the evaluation  of the  lowest (\it Bogomol'nyi\rm)  bound $E_{min}$, (\ref{Emin}), of the "(YM- Higgs)" energy, associated  with  vacuum  \linebreak monopole solutions (this bound depends  on the Higgs mass $m/\sqrt{\lambda}$),
allowed us to specify
 monopole configurations of the Minkowskian YM vacuum as
BPS or Wu-Yang  monopoles
(obtained as  infinite spatial volume limits of  BPS monopoles).\par
BPS monopoles (\ref{sc monopol}), (\ref{YM monopol}) (in the zero topological sector of the Minkowskian YM model \cite{LP1, Pervush2,David2,David1}) are regular in a finite spatial volume (although we cannot treat  Higgs BPS  monopole solutions (\ref{sc monopol}) as completely  regular ones: they diverge at the spatial infinity; on the
other hand 
it is a large advantage, as far as we may treat Higgs vacuum solutions 
(\ref{sc monopol}) as  \it a singular  Bose condensate \rm \cite{Pervush1}, involving the appearance of  vacuum "electric"   monopoles (\ref{el.m})). 
\par 
We have described
the topological degeneration  of initial data for  monopole solutions belonging to the zero topological sector of the Minkowskian YM theory \cite{LP1, Pervush2,David2,David1}. This topological degeneration
comes to Gribov copies of the covariant Coulomb gauge (\ref{Aparallel}), treated
 as  zero initial data for the YM Gauss law constraint (\ref{Gauss}) (this implies  the absence of longitudinal
YM fields in the initial time instant $t_0$). \par
As far as YM fields are massless, the imposition of the
transverse Coulomb gauge implies (see, e.g.,  \cite{Rohrlich}) the
removal of temporal components of YM fields. We should necessarily remove these  temporal YM components as long as they are non-dynamical degrees of freedom the existence of which contradicts the Heisenberg uncertainly principle \cite{Pervush2,Dir}. \par
The removal  a la Dirac \cite{Dir} of temporal components of YM fields turns latter ones into topological  Dirac variables (\ref{degeneration1}) \cite{LP1,David2, David1}, transverse and physical.  They are stationary and 
depend only on "large" (with topological numbers  $n\neq 0$) matrices $v^{(n)}({\bf x})$ of the  Gribov topological degeneration (belonging to the nontrivial  $U(1)\to SU(2)$ embedding).
 \par
The Coulomb gauge (\ref{Aparallel}) is not defined in the unique way.
It is a purely non-Abelian effect \cite{Al.S.,Baal} called the  Gribov
 ambiguity. To find  the Gribov  ambiguity in the choice of the Coulomb gauge (\ref{Aparallel}), we should solve the Gribov ambiguity   equation (\ref{Gribov.eq}) of the second order.\par
Considering
 the  Gribov topological degeneration  in the given initial time instant $t_0$
ensures  that  Dirac dressing matrices $U^D$ \cite{LP1}, (\ref{UD}), remain  trivial  in this  time instant, in despite of the Gribov ambiguity.  Thus the Gribov ambiguity does not affect the nature of the Gribov topological degeneration. \par
In turns, this allows us to solve the Gribov ambiguity problem as the  Cauchy  task (\ref{Gribov.eq}) with the  initial conditions (\ref{Gauss2}) (that is mathematically equivalent to (\ref{init})) and  (\ref{Aparallel}) in the time instant $t_0$, i.e. on  the fixed space-like surface ${\cal H} (t_0)$ in the  Minkowski space-time.
This means that  we 
find topologically degenerated YM fields satisfying the Coulomb gauge (\ref{Aparallel}),   (\ref{transv}) (transverse and physical topological Dirac variables (\ref{degeneration1})) in the class of vacuum BPS monopoles and perturbation excitations over this monopole vacuum\rm. \par 
The  Gribov equation (\ref{Gribov.eq}) describes the  nontrivial cohomological
structure of  YM fields in the Minkowski space.
It turns out that there exists a one-to-one
correspondence between the set of cohomologies classes of YM
fields and the one of  Gribov copies of the Coulomb gauge (\ref {Aparallel}).
This cohomological structure corresponds to
the elements of the holonomies group $H$ constructed over  transverse YM
fields belonging to the nontrivial $U(1)\to SU(2)$ embedding. \par
The unit element of the holonomies group $H$ (the whole this group is isomorphic to the residual gauge symmetry group $U(1)$) is degenerated
with respect to the class of exact 1-forms (with  zero
topological charges) induced by the Coulomb gauge (\ref {Aparallel}) 
and  Bogomol'nyi equation (\ref{Bog}). \par
As a result, we may pick out the restricted holonomies subgroup $\Phi^0$ in the holonomies group $H$ isomorphic to the  subgroup of "small" $U(1)$ gauge transformations in the residual $U(1)$ gauge group. We may treat the above nontrivial cohomological
structure of  YM fields as the solution of the Cauchy  task (\ref{Gribov.eq}), (\ref{Aparallel}), (\ref{init}).
\par
  YM fields  are considered \cite{LP1} as  sums of
vacuum   fields (BPS monopoles) and  weak perturbation excitations over this vacuum
(multipoles).
We suppose that  multipoles possess the same topological numbers
 that appropriate BPS monopoles. \par
The important point of our investigations is
that the  square of the  vacuum expectation
value of the "magnetic" tension, $<B^2>$, is different from zero in the Minkowskian YM model \cite{LP1, Pervush2,David2,David1}. This non-zero
"magnetic" tension is an important distinction of the Minkowskian  YM theory \cite{LP1, Pervush2,David2,David1} from the Euclidian one \cite{Bel}, where the zero asymptotic of the 
"magnetic" tension at the infinity \cite{Cheng} ensures the existence of instanton YM solutions.  
\par
We have shown that there  exists the continuous topological variable $N(t)$
 specifying the zero mode of the Gauss law constraint (\ref{Gauss})  and depending on
the time $t$; it plays the
role of the non-integer degree of the map. \par
The introduction of the  topological variable $N(t)$
results the appearance of vacuum "electric"  monopoles   (\ref{el.m})  in the Minkowskian  YM theory \cite{LP1, Pervush2,David2,David1}, induced by
temporal components of YM fields,  (\ref {sol.zero}), complementary to equal to 
 zero (due the above resolving of the Gauss law constraint in terms of 
the Coulomb gauge (\ref {Aparallel})) ones in the "pure" YM  theory. 
It is also the purely 
Higgs vacuum effect. \par
The calculations associated with the topological variable $N(t)$
and vacuum "electric"  monopoles   (\ref{el.m}) involve the action for the free rotator (\ref{rot}):
with the rotary momentum $I$ depending  on $<B^2>$ and
  real spectrum of momentum\rm. This
spectrum  describes the  collective solid potential rotation of the  (YM-Higgs) vacuum. 
\par
We have constructed the complete  Hamiltonian (\ref{Hamilton}) \cite{LP1} of the 
Minkowskian  YM vacuum \cite{LP1, Pervush2,David2,David1}. It  consists of  two parts: "electric" and " magnetic"
ones.  This Hamiltonian is explicitly Poincare (in particular, CP) invariant, unlike the well-known
$\theta$-term  \cite{Cheng} in the instanton YM Euclidian theory \cite{Bel}. Thus  the CP problem
may be solved in the  Minkowskian  YM theory \cite{LP1, Pervush2,David2,David1} in which vacuum YM and "electric" 
monopole modes
appear.\par
The  Coulomb gauge (\ref {Aparallel}), (\ref{transv}) allows us to solve the Gauss  law constraint (\ref{Gauss})
in terms of its zero mode  (\ref{Summ}), i.e.  vacuum "electric"  monopoles.  As a result, the transverse "electric" 
 tension (\ref{pop.napr}) appears. It is again the purely 
Higgs effect. \par
 On the other hand, the analysis of the  constraint-shell action  (\ref{int.sum}) allowed us to draw the following very important conclusion: that vacuum
scalar BPS monopoles (and  "electric"  ones together with) disappear from the excitations spectrum in the 
infinite spatial volume limit \rm $V\to\infty$\rm. Nevertheless,   Higgs BPS monopoles
 leave their trace 
in the unconstrained system (\ref{unconstr}) of local excitations
as a weak "electric" 
 tension $\tilde E$: a quantum fluctuation over the  "electric"  monopole.  \par
This field plays a crucial role in the definition (\ref{total current})  \cite{David1} of the total 
current: the sum of the non-Abelian and fermionic components. We treat this current 
 as an excitation over the physical (Minkowskian) YM vacuum. \par
The total  current (\ref{total current}) satisfies Eq. (\ref{cur-t})  of the second order, depending on Wu-Yang (BPS) monopoles. \par
We have found the Green function of  this equation
as a composition of  two potentials: the Coulomb type potential (\ref{Cp})  and  non-linear rising one ("the golden section potential")  (\ref{ris1}). \par 
The latter  potential implies the modified (in the infrared momenta region) gluonic  propagator,
alike  (\ref{modif}) \cite{Bogolubskaja,Werner}, and causes the hadronization of quarks. \par
The considered  nontrivial topological structure of the vacuum in the Minkowskian 
 YM theory \cite{LP1, Pervush2,David2,David1}  may be in  other non-Abelian Minkowskian theories.
For  instance,
 there is the spontaneous $SU(3)_{col}\to SU(2)$
breakdown with the antisymmetric choice of the Gell-Mann matrices:
$\lambda_2, \lambda_5, \lambda_7$, involving vacuum Wu-Yang monopoles (see 
Eqs. (3.24) -(3.25) in \cite {David2}).
The essential point of the theory \cite {David2} was \it the mix of
  world and  group indices \rm at
constructing   vacuum Wu-Yang monopoles\rm. \par
One may consider the behaviour of
quarks  in the Wu-Yang monopole background and  compute the Green function of  a quark (see (4.9)- (4.13) in \cite {David2}). \par
Arguments in  favour  of the  non-Abelian Minkowskian theory \cite{LP1, Pervush2,David2,David1}  
are also the 
 additional mass of the $\eta_0$-meson, $\triangle m_\eta$, and 
the  infrared topological confinement \cite{Pervush2, Pervush3,Ilieva,Nguyen,Azimov}.\par
So, speaking about the calculation about the additional mass of the 
$\eta_0$-meson in the non-Abelian Minkowskian model \cite{LP1, Pervush2,David2,David1}  (this may be considered  as the specific way to solve the $U(1)$-problem in the Minkowski space\rm), it is worth to note the crucial role of (topologically trivial) Higgs BPS monopole modes (\ref{sc monopol}) in forming the mesonic  mass $\triangle m_\eta$.  \par
Really, indeed
$$\triangle m_\eta \sim 1/\sqrt {I_{QCD}}.$$
On the other hand, the rotary momentum $ I_{QCD}$, (\ref{I}), of the non-Abelian Minkowskian vacuum is specifies by topologically trivial Higgs BPS monopole modes (\ref{sc monopol}). \par
Moreover, this implies that the mesonic  mass $\triangle m_\eta$ is directly proportional to  $<B^2>\neq 0$, the nonzero vacuum expectation value of the "magnetic" tension squared inherent in the non-Abelian Minkowskian model \cite{LP1, Pervush2,David2,David1} and induced by the Bogomol'nyi equation   (\ref{Bog}). 
\par
Discussing the  infrared topological confinement \cite{Pervush2, Pervush3,Ilieva,Nguyen,Azimov} in the non-Abelian Minkowskian model \cite{LP1, Pervush2,David2,David1}, we have demonstrated that "large" Gribov topological multipliers $v^{(n)}({\bf x})$ ($n\neq 0$) disappear \cite{Azimov} in quark Green functions in all the orders of the perturbations theory in the limit of small transferred  momenta. \par 
Such effect is reached by virtue of the infrared spatial asymptotic (\ref{bondari}) \cite{Azimov} of "large" Gribov topological multipliers $v^{(n)}({\bf x})$. In turn, this asymptotic implies the normalization
(\ref{fas.ysl-e}) for the Euler angles $\phi_i$ ($i=1,2,3$).\par 
We have demonstrated, recalling the arguments \cite{Azimov},  that  the infrared topological
confinement  implies the colour confinement in Minkowskian QCD \cite{LP1, Pervush2,David2,David1}. 
This means that
only the colourless ("hadronic") states (Feynman partons)  may be treated as  physical states in that QCD. Minkowskian QCD \cite{LP1, Pervush2,David2,David1} is a unitary theory with respect to these states.
\par
The criterion of the colours confinement in Minkowskian QCD \cite{LP1, Pervush2,David2,David1}, involving 
 the  Coulomb  gauge (\ref{Aparallel}),
is the 
existence of the nontrivial restricted holonomies group $\Phi ^0$ 
constructed on the transverse YM
fields of the zero topological sector belonging to the $U(1)\to SU(2)_{col}$ embedding. In addition these fields would satisfy the boundary condition (\ref{bondari}) \cite{Azimov} to ensure the infrared topological (and colours) confinement in Minkowskian QCD \cite{LP1, Pervush2,David2,David1}.\par
The said allowed us to describe the colour confinement in Minkowskian QCD \cite{LP1, Pervush2,David2,David1} in terms of the mixed task, in the space-like  surface ${\cal H}(t_0)$ in the  Minkowski space-time, to the Gribov ambiguity equation (\ref{Gribov.eq}). \par
This mixed task to the Gribov ambiguity equation (\ref{Gribov.eq}) comes to the Cauchy task  (\ref{Gribov.eq}), (\ref{init}),  (\ref{Aparallel}) (responsible for the topological degeneration of initial YM data in the Minkowskian theory \cite{LP1, Pervush2,David2,David1}) supplemented by the boundary condition (\ref{bondari}) \cite{Azimov} to Gribov topological multipliers $v^{(n)}({\bf x})$  at the spatial infinity. \par
The Lorentz covariance of the considered theory may be carried out via  Lorentz rotations of the
time axis $l_\mu^{(0)}$ along the complete momentum of each of the
physical states, i.e. by  transitions to  reference frames 
where the initial data and  spectra of these states are measured
\cite {David2}.
\par 
All these "smiles" of the Higgs scalar field disappear as we replace
the fundamental Dirac variables \cite{Dir,Nguyen2,FJ} and change
the gauge of their physical sources in order to obtain
the conventional Faddeev-Popov integral \cite{Fadd3} as a realization of
the Feynman heuristic quantization \cite{AGP}.
We have made sure in this with the convincing example of the mixed task (\ref{Gribov.eq}), (\ref{init}),  (\ref{Aparallel}), (\ref{bondari}) to the Gribov ambiguity equation (\ref{Gribov.eq}) explained the colour confinement in Minkowskian non-Abelian theory \cite{LP1, Pervush2,David2,David1}. Going over from the  Coulomb transverse  gauge (\ref{Aparallel}) to an arbitrary another one destroys  this mixed task.
\par
Such change removes, in general, all the time axes of the physical states, all the initial
data, with their degeneration and destructive interference, and all
monopole effects, including instantaneous interactions forming
non-local bound states of the types of atoms in QED or hadrons in QCD. \par
In other words, the "smiles" of the Higgs field show us
the limitedness of the FP heuristic path integral.
 The generalization of the
Faddeev theorem of equivalence \cite{Fadd1} (that is valid, indeed, only for  local scattering
processes) to the region of  non-local processes removes
both the initial data
and the Laplace possibility of explaining (by these data) the non-local physical
effects of the type of hadronization and confinement  in
this world.
\section*{Acknowledgments.}
I should like to thank Prof. V. N. Pervushin, my co-author by series of recent publications, for his useful advices and recommendations during the preparation of the present paper.  \par
I am grateful to Prof. D. Ebert, R. Jackiw, V. G. Kadyshevsky, F. Lenz, M. M$\ddot u$ller-Preusker, O. S. Parasiuk, Yu. P. Stepanovsky, L. Susskind,  V. I. Tkach,  Dr. \linebreak D. Antonov,  E. -M. Ilgenfitz  for  fruitful discussions and critical remarks concerning the present investigations.  \par 
I should like, on behalf of the authors collective (L. L, V.P.), to thank Dr. Yu. Reznik for his interest to   our previous publication \cite{LP1} and his remark about this work.

\enddocument